\documentclass[journal]{IEEEtran}

\usepackage[OT1]{fontenc}
\usepackage[numbers,sort&compress]{natbib}
\usepackage[cmex10]{amsmath}
\usepackage{amssymb}
\usepackage{bm}
\usepackage{braket}
\usepackage{graphicx}
\usepackage{color}
\usepackage{subfigure}
\usepackage{tabularx}
\usepackage{arydshln}
\usepackage{mathtools}
\usepackage{multirow}
\usepackage{algorithm}
\usepackage{algpseudocode}
\usepackage{url}
\usepackage{listings}
\usepackage{comment}
\usepackage{hyperref}
\usepackage[acronym,shortcuts]{glossaries}

\def\Y{\mathrm{Y}}

\def\H{\mathrm{H}}

\def\X{\mathrm{X}}

\def\A{\mathrm{A}}

\def\R{\mathrm{R}}
\def\b{\mathrm{b}}

\def\vec{\mathrm{vec}}

\def\U{\mathrm{U}}

\def\x{\mathrm{x}}
\def\G{\mathrm{G}}
\def\D{\mathrm{D}}

\newacronym{QAP}{QAP}{quadratic assignment problem}
\newacronym{GAS}{GAS}{Grover adaptive search}
\newacronym{IQFT}{IQFT}{inverse quantum Fourier transform}
\newacronym{QUBO}{QUBO}{quadratic unconstrained binary optimization}
\newacronym{HUBO}{HUBO}{higher-order unconstrained binary optimization}
\newacronym{PPO}{PPO}{permutation preparation oparator}

\newcommand{\plainfootmark}{\refstepcounter{footnote}\textsuperscript{\thefootnote}}
\newcommand{\plainfoottext}[1]{\footnotetext{#1}}

\begin{document}
\title{Grover Adaptive Search for the Higher-Order Formulation of Quadratic Assignment Problems}

\author{Taku~Mikuriya, Shintaro~Fujiwara, Kein~Yukiyoshi,~\IEEEmembership{Graduate Student~Member,~IEEE}, Giuseppe~Thadeu~Freitas~de~Abreu, and Naoki~Ishikawa,~\IEEEmembership{Senior~Member,~IEEE}.
\thanks{T.~Mikuriya, S.~Fujiwara, K.~Yukiyoshi and N.~Ishikawa are with the Faculty of Engineering, Yokohama National University, 240-8501 Kanagawa, Japan (e-mail: ishikawa-naoki-fr@ynu.ac.jp). G.~T.~F.~Abreu are with the School of Computer Science and Engineering, Constructor University, Campus Ring 1, 28759, Bremen, Germany. This research was partially supported by JST CRONOS Japan Grant Number JPMJCS24N1.}
\vspace{-4ex}}

\markboth{\today}
{Shell \MakeLowercase{\textit{et al.}}: Bare Demo of IEEEtran.cls for Journals}
\maketitle

\begin{abstract}
We demonstrate that the search space of the quadratic assignment problem (QAP), known as an NP-hard combinatorial optimization problem, can be reduced using Grover adaptive search (GAS) with permutation preparation operator (PPO). To that end, we first revise the traditional quadratic unconstrained binary optimization (QUBO) formulation of the QAP into a higher-order unconstrained binary optimization (HUBO) formulation, introducing a binary encoding method. Algebraic analyses in terms of the number of qubits, quantum gates, circuit depth, and query complexity are performed, which indicate that our proposed approach significantly reduces the search space size, improving convergence performance to the optimal solution compared to the conventional one. Furthermore, although the PPO for HUBO has a greater circuit depth than the PPO for QUBO, when the analysis is extended to the entire state preparation operator, both HUBO and QUBO exhibit comparable depths. Therefore, owing to its smaller number of variables, HUBO can be concluded to be more effective.
\end{abstract}


\IEEEpeerreviewmaketitle

\section{Introduction}

The \ac{QAP} was formulated in $1957$ by Koopmans and Beckmann as a mathematical model to represent interactions arising in various allocation tasks in economic activities \cite{koopmans1957assignment}.
Variations of the \ac{QAP} appear in a wide range of real-world problems such as facility location, backboard wiring, scheduling, and the allocation of quasi-Gray codewords to symbols of multidimensional constellations in communications systems \cite{colman2011quasigray}.

Known to be of nondeterministic polynomial-time (NP) complexity, usual methods to solve this problem, such as migrating birds optimization \cite{duman2012migrating}, simulated annealing \cite{wilhelm1987solving}, tabu search \cite{rego2010ejection}, ant colony optimization \cite{talbi2001parallel}, and breakout local search \cite{benlic2013breakout}, are of a heuristic nature. It has been shown that finding even approximate solutions for the \ac{QAP} in polynomial time becomes infeasible as the problem size increases even under such heuristic methods \cite{sahni1976pcomplete}.

In turn, quantum-related technologies to address NP-hard combinatorial optimization problems can potentially be used to circumvent the latter challenge.
In particular, coherent Ising machines \cite{yamamoto2017coherent} and quantum annealing \cite{kadowaki1998quantum} are designed for \ac{QUBO} problems.
For \ac{HUBO} problems, however, additional auxiliary variables are needed, which expands the search space and increases the complexity of finding solutions.

Quantum gate-based algorithms such as the quantum approximate optimization algorithm (QAOA) \cite{campbell2021qaoa} and the \ac{GAS} algorithm \cite{gilliam2021grover}, on the other hand, can handle objective functions of third or higher orders without such issues.
In particular, since GAS is an exhaustive search algorithm, it is theoretically guaranteed to obtain the optimal solution.
To elaborate, \ac{GAS} provides a quadratic speedup for a \ac{QUBO} or \ac{HUBO} problem over a classical exhaustive search.
Specifically, for $n$ binary variables, a classical exhaustive search requires a query complexity of $O(2^n)$, whereas \ac{GAS} reduces the latter requirement to $O(\sqrt{2^n})$.

\begin{table}[t]
\centering
\label{table:sym}
\caption{List of important mathematical symbols.}
\vspace{-2ex}
\begin{tabular}{lll}
\hline\\[-2ex]
$\mathbb{B}$ & & Binary numbers \\
$\mathbb{N}$ & & Natural numbers\\
$\mathbb{R}$ & & Real numbers \\
$\mathbb{C}$ & & Complex numbers\\
$\mathbb{Z}$ & & Integers \\
$x$ & $\in \mathbb{B}$ & Binary variable \\
$\mathrm{x}$ & & Binary variables \\
$E(\cdot)$ & $\in \mathbb{R}$ & Objective function\\
$\mathrm{j}$ & $\in \mathbb{C}$ & Imaginary number\\
$n$ & $\in \mathbb {Z}$ & Number of binary variables\\
$m$ & $\in \mathbb{Z}$ & Number of qubits to encode $E(\cdot)$\\
$y$ & $\in \mathbb{Z}$ & Threshold for $E(\cdot)$ \\
$N$& $\in\mathbb{N}$&Problem size\\
$\mathrm{F}$ & $\in\mathbb{R}^{N\times N}$ & Flow matrix\\
$\mathrm{D}$& $\in\mathbb{R}^{N\times N}$ & Distance matrix\\
$\mathrm{X}$& $\in\mathbb{B}^{N\times N}$ &Solution matrix of QUBO\\
$\mathrm{Y}$& $\in\mathbb{B}^{N\times N}$ &Solution matrix of HUBO\\
$\mathrm{P}$& $\in\mathcal{P}$& Permutation matrix\\
$\mathcal{P}_n$& &Set of all permutation matrices of size $n$\\
$\ket{W_n}$ &$\in \mathbb{C}^{2^n}$ & W state of size $n$\\
$\ket{S_M}$ &$\in \mathbb{C}^{2^n}$ & Shukla state for integer $M$\\[1ex]
\hline
\end{tabular}
\vspace{-1ex}
\end{table}

However, it is typical in current approaches to apply Hadamard gates to all qubits during the initial state preparation of \ac{GAS}  \cite{ishikawa2021quantum, sano2024accelerating, sano2023qubit, yukiyoshi2022quantum, norimoto2023quantum, huang2023optimization}, resulting in a superposition that includes infeasible solutions, increasing the query complexity of the algorithm.

In order to address this issue, a method for modifying the initial state of \ac{GAS} has been proposed \cite{yukiyoshi2024quantum}.
For a \ac{QAP} of size $N$, which has $N!$ feasible solutions, a quantum circuit that generates an equal superposition of these solutions has been extensively studied \cite{fefferman2015power,bartschi2020grover,matsuo2023enhancing,liu2025lowdepth}.

Against this background, in this paper, we propose two formulations to solve the \ac{QAP}: 1) a \ac{QUBO} formulation with \ac{PPO} \cite{matsuo2023enhancing}, which generates a uniform superposition of all permutation patterns; and 2) a \ac{HUBO} formulation using a new \ac{PPO}.
These two proposed formulations aim to efficiently solve the \ac{QAP} using the \ac{GAS} algorithm.
Both of these approaches demonstrate that the \ac{QAP} can be solved more efficiently than by using the conventional Hadamard transformation, and we emphasize in particular that the \ac{HUBO} formulation is more effective due to its smaller qubit requirement. For example, \cite{matsuo2023enhancing} proposed a construction that uses $N^2$ qubits without any ancilla qubits and achieves a depth of $O(N^2)$, while \cite{liu2025lowdepth} demonstrated a construction that employs $N^2$ qubits together with $O(N \log N)$ ancilla qubits and attains logarithmic-order depth. Thus, these approaches yield widely varying advantages. It should be noted that among these, only \cite{matsuo2023enhancing} provides details on the specific construction of the quantum circuit.

All in all, the contributions of this paper are as follows.
\begin{enumerate}
\item The \ac{QUBO} formulation of the \ac{QAP} is reworked by preparing the initial state using \ac{PPO} for QUBO \cite{matsuo2023enhancing}, thereby reducing the corresponding search space.

\item A \ac{HUBO} formulation of the \ac{QAP} is proposed, in which a binary encoding is devised to reduce the required number of qubits.
\item \ac{PPO} for \ac{HUBO} is proposed. To the best of our knowledge, no prior work has proposed a higher-order formulation of the \ac{QAP} together with the corresponding PPO, and this manuscript represents the first attempt in this direction.

\item Algebraic analyses of the number of terms in the objective function, qubits, and quantum gates, as well as circuit depth and query complexity are conducted. Both \ac{QUBO} and \ac{HUBO} outperform the conventional Hadamard transformation in terms of query complexity.
\end{enumerate}

The remainder of the article is organized as follows.
In Section \ref{sec:preliminaries}, an overview of the \ac{QAP} and its conventional \ac{QUBO} formulation, as well as a brief description of the \ac{GAS} algorithm is provided.
Then, the \ac{GAS} algorithm is revised in Section \ref{sec:Proposed QUBO} by using the \ac{PPO} for \ac{QUBO}.
In Section \ref{sec:Proposed HUBO}, the proposed \ac{HUBO} formulation relying on the new \ac{PPO} for \ac{HUBO} is described. The corresponding \ac{GAS} algorithm is also discussed.
Algebraic and numerical analyses of the proposed formulations are then carried out in Section \ref{sec:analysis}.
Finally, Section \ref{sec:conc} concludes the paper.

A list of important mathematical symbols used throughout the article is presented in Table \ref{table:sym}.
Hereafter, italicized symbols will consistently be used to represent scalar values, while roman symbols represent vectors and matrices.

\vspace{-1ex}
\section{Preliminaries}
\label{sec:preliminaries}

\subsection{The Quadratic Assignment Problem}
\label{sec:qap}

The \ac{QAP} \cite{cela2013quadratic} is a fundamental combinatorial optimization problem, which can be explained in terms of its original facility location context arising in Economics, and amounts to finding the optimal assignment of $N$ facilities to $N$ locations, such that the total cost of all economic transactions among the facilities is minimized, where the cost of a transaction between any two facilities is given by the product of the flow and the distance between their locations.

Denoting the flow between facilities $i$ and $j$ by $f_{ij}$, and the distance between locations $k$ and $\ell$ by $d_{k\ell}$, the Koopmans-Beckmann formulation of the \ac{QAP} \cite{koopmans1957assignment} is given by
\begin{align}  \underset{\phi\in\mathfrak{S}_N}{\text{minimize}}\sum_{i=0}^{N-1}\sum_{j=0}^{N-1} f_{ij}d_{\phi(i)\phi(j)},
\label{eq:KB-QAP}
\end{align}
where $\phi$ denotes a permutation of the set $\mathcal{N}=\{0,\cdots,N-1\}$, such that $f_{ij}d_{\phi(i)\phi(j)}$ represents the cost of assigning facilities $i$ and $j$ to locations $\phi(i)$ and $\phi(j)$, respectively, while $\mathfrak{S}_N$ is the set of all permutations $\phi$ of $\mathcal{N}$.

Note that the latter notation $\mathfrak{S}_N$ implies that each facility is assigned to a single location, and that each location is occupied by a single facility.  
In addition, a given permutation $\phi$ can be represented by a permutation matrix $\mathrm{P}=(p_{ij})\in \mathcal{P}_N\in\mathbb{B}^{N\times N}$ defined as  
\begin{align}
p_{ij}=\left\{
\begin{array}{cc}
1  & \mathrm{if\;}\phi(i)=j, \\
0  & \mathrm{otherwise,}
\end{array}
\right.
\end{align}
corresponding to $\phi = \mathrm{P}[0,\dots,N-1]^\text{T}$, with the set of permutation matrices of size $N$ denoted by $\mathcal{P}_N$, whose cardinality is $|\mathcal{P}_N| = N!$.

The relationship between a certain permutation $\phi\in \mathfrak{S}_N$ and the corresponding permutation matrix $\mathrm{P}\in\mathcal{P}_N$ is given by
\begin{align}
\left(\mathrm{P}\mathrm{D}\mathrm{P}^{\mathrm{T}}\right)_{ij}=d_{\phi(i)\phi(j)},
\end{align}
where $\mathrm{D} \triangleq (d_{k\ell})\in\mathbb{R}^{N\times N}$ is the distance matrix.

Defining the flow matrix as $\mathrm{F} \triangleq (f_{ij})\in\mathbb{R}^{N\times N}$, and the inner product between two generic matrices $\mathrm{A}=(a_{ij})\in\mathbb{R}^{N\times N}$ and $\mathrm{B}=(b_{ij})\in\mathbb{R}^{N\times N}$ as
\begin{align}
\langle\mathrm{A}, \mathrm{B}\rangle\triangleq\sum_{i=0}^{N-1}\sum_{j=0}^{N-1} a_{ij}b_{ij},
\end{align}
the Koopmans-Beckmann's \ac{QAP} formulation can be concisely described by
\begin{equation}
\underset{\mathrm{P}\in\mathcal{P}_N}{\text{minimize}} \quad \langle\mathrm{F},\mathrm{P}\mathrm{D}\mathrm{P}^{\mathrm{T}}\rangle.\\
\label{eq:QAP}
\end{equation}

\subsection{QUBO Formulation of the QAP}
\label{subsec:Conv_BO}

Defining the binary space $\mathbb{B}\triangleq \{0,1\}$, and corresponding generic binary matrices $\mathrm{X}\in\mathbb{B}^{N\times N}$, the problem described by equation \eqref{eq:QAP} can be reformulated as the binary optimization problem \cite{glover2019quantum}
\begin{align}
\underset{\mathrm{X}\in\mathbb{B}^{N\times N}}{\text{minimize}} \quad & \left< \mathrm{F}, \mathrm{X} \mathrm{D} \mathrm{X}^{\mathrm{T}} \right>\\
\mathrm{s.t.}
\quad & \sum_{i=0}^{N-1} x_{ij} = 1 \quad (j=0,\cdots,N-1)\nonumber \\
\quad & \sum_{j=0}^{N-1} x_{ij} = 1 \quad (i=0,\cdots,N-1)\nonumber \\
\quad & x_{ij} \in \mathbb{B}.\nonumber
\label{BO_obj}
\end{align}

Notice that the constraints in the latter problem ensure that each column/row of the solution $\mathrm{X}$ has a single non-zero entry, which in this context, where $i$ and $j$ denote the indices of facilities and locations, respectively, implies that the binary variable $x_{ij}$ is defined as
\begin{align}
x_{ij}=\left\{
\begin{array}{cl}
1 &  \mathrm{if \;facility\;} i \mathrm{\;is\;assigned\;to\;location\;} j, \\
0 &  \mathrm{otherwise.}
\end{array}
\right.
\end{align}

Furthermore, observe that the number of binary variables $x_{ij}$ required to describe a solution is given by $N^2$.
However, since there are $N!$ possible solutions, the minimum number of binary variables required to represent all solutions can be estimated, using Stirling's approximation \cite{george1990anew}, to be
\begin{equation}
\begin{split}
\log_2(N!)&=O(N\log N),
\end{split}
\label{eq:log2Nf}
\end{equation}
which indicates that the binary formulation given in equation \eqref{BO_obj} is not optimal in terms of the number of binary variables, leading to increased search space size.

Finally, note that by moving the constraints in the quadratic binary problem \eqref{BO_obj} into the objective as regularization terms, a \ac{QUBO} formulation of the problem can be obtained, namely
\vspace{-1ex}
\begin{align}
\label{eq:QUBO_obj}
\underset{\mathrm{X}\in\mathbb{B}^{N\times N}}{\text{minimize}}\left< \mathrm{F}, \mathrm{X} \mathrm{D} \mathrm{X}^{\mathrm{T}} \right> &+\lambda_1\sum_{i=0}^{N-1}\bigg(\sum_{j=0}^{N-1} x_{ij}-1\bigg)^{2}\\
&+\lambda_2\sum_{j=0}^{N-1}\bigg(\sum_{i=0}^{N-1} x_{ij}-1\bigg)^{2},\nonumber
\vspace{0.5ex}
\end{align}
where $\lambda_1$ and $\lambda_2$ are penalty coefficients and we implicitly define, for future reference, the objective function $E(\x)$, with $\x \triangleq \mathrm{vec}(\mathrm{X}) \in \mathbb{B}^{N^2}$, where $\vec(\cdot)$ denotes the row-major vectorization.

\vspace{-2ex}
\subsection{The Quantum Adaptive Search Algorithm}
\label{subsec:GAS}

For unstractured search, Grover's search algorithm has been proven to offer a quadratic speedup compared to the classical exhaustive search approach \cite{grover1996fast}.
The technique has been extended to solve the minimum value search problem \cite{durr1999quantum}, and to cases where the number of solutions is unknown \cite{boyer1998tight}.
Although the oracle that flips only the phase of the desired state has been treated as an idealized component in earlier/traditional studies, this issue has been solved by Gilliam et al. \cite{gilliam2021grover}, in which a well-defined quantum circuit corresponding to the \ac{QUBO} or \ac{HUBO} formulation of the problem is efficiently constructed.

The quantum circuit constructed in \cite{gilliam2021grover} is composed of $n + m$ qubits, where $n$ is the number of binary variables and $m$ is the number of qubits representing the value of the objective function $E(\x)$ used for the search, with the latter satisfying
\begin{equation}
\label{eq:constraint_of_m}
-2^{m-1} \leq \min_{\x}[E(\x)] \leq \max_{\x} [E(\x)]<2^{m-1}.
\end{equation}

In \ac{GAS}, an iteration index $i$ is initialized to $0$, and the objective function value $y_0=E(\x_0)$ corresponding to a random initial solution $\x_0$ is used as a threshold for sampling solutions whose objective function values are less than $y_0$.
Then, measurements through a quantum circuit $\G^{L_i}\A_{y_i}\ket{0}_{n+m}$ are repeated until a termination condition is satisfied.

\begin{algorithm}[H]
\vspace{1ex}
\caption{GAS for real-valued coefficients  \cite{gilliam2021grover,norimoto2023quantum}.\label{alg:real-gas}}
\begin{algorithmic}[1]
\renewcommand{\algorithmicrequire}{\textbf{Input:}}
\renewcommand{\algorithmicensure}{\textbf{Output:}}
\Require $E:\mathbb{B}^{n}\rightarrow\mathbb{Z}, \lambda=8/7$
\Ensure $\x$
\State {Uniformly sample $\x_0 \in \mathbb{B}^n$ and set $y_0=E(\x_0)$}.
\State {Set $k = 1$ and $i = 0$}.
\Repeat
\State{Randomly select $L_i$ from a set $\{0, 1, ..., \lceil k-1 \rceil$\}}.
\State{Evaluate $\G^{L_i} \A_{y_i} \Ket{0}_{n + m}$ to obtain $\x$ and $y\! =\! E(\x)$}.
\Statex\Comment{This is an additional step to support real coefficients.}
\hspace{\algorithmicindent}\If{$y<y_i$}
\State\hspace{\algorithmicindent}{$\x_{i+1}=\x, y_{i+1}=y,$ and $k=1$}.
\Else
\State\hspace{\algorithmicindent}{$\x_{i+1}=\x_i, y_{i+1}=y_i,$ and 
\Statex \hspace{\algorithmicindent}\hspace{\algorithmicindent}\hspace{\algorithmicindent}$k=\min{\{\lambda k,\sqrt{2^n}}\}$}.
\EndIf
\State{$i=i+1$}.
\Until{a termination condition is met}.
\end{algorithmic}
\end{algorithm}

Here, $L_i$ iterations of the Grover operator are applied, where $L_i$ is chosen from a uniform distribution over a half-open interval $[0,k)$, and $k$ is updated with $\min\{\lambda k, \sqrt{2^n}\}$ if the sampled objective function value exceeds the threshold, and initialized to $1$ if it falls below it.
The procedure is summarized in Algorithm \ref{alg:real-gas}.
In turn, the construction of the actual state preparation operator $\A_y$ of the quantum circuit $\G^{L_i}\A_{y_i}\ket{0}_{n+m}$ can be summarized in the following three steps.

\subsubsection{Initial State Preparation} 
\label{subsubsec: initial state preparation of GAS}
First, the initial state corresponding to the solution space $\mathcal{X}$ of the problem is prepared. A commonly employed method for generating the initial state is the Hadamard transformation, which applies Hadamard gates to all $n$ qubits \cite{ishikawa2021quantum, norimoto2023quantum, sano2024accelerating}, thereby creating a uniform superposition over the entire solution space representable by $n$ qubits.

If the solution space is subject to constraints, it is possible to improve the efficiency of the search by adjusting the initial state accordingly. For instance, under a one-hot constraint in which exactly one of the $n$ variables is $1$ and all others are $0$, the W state $\ket{W_n}$ defined by the following expression can be used \cite{cruz2019efficient}:
\begin{equation}
\ket{W_n} = \frac{1}{\sqrt{n}}\sum_{\substack{\x\in\{0,1\}^n \\ H_w(\x)=1}} \ket{\x}_n,
\end{equation}
where $H_w(\x)$ is the Hamming weight of the bit string $\x$.

For a more generalized solution space where exactly $k$ out of $n$ variables are $1$ and the rest are $0$, the Dicke state $\ket{D^n_k}$ defined as follows can be employed \cite{bartschi2022shortdeptha}.
\begin{equation}
\ket{D^n_k} = \frac{1}{\sqrt{\binom{n}{k}}}\sum_{\substack{\x\in\{0,1\}^n \\ H_w(\x)=k}} \ket{\x}_n
\end{equation}

A method utilizing the Dicke state as the initial state for GAS has already been proposed \cite{yukiyoshi2024quantum}.
By applying a unitary operator that prepares an initial state spanning the solution space $\mathcal{X}$ for the $n$ qubits corresponding to the variables, and further applying the Hadamard gate $\H$ to an $m$-bit register to represent arbitrary objective function values, the state evolves as 
\begin{equation}
\ket{0}_{n+m}\xrightarrow{} \frac{1}{\sqrt{2^m|\mathcal{X}|}}\sum_{\x\in\mathcal{X}}\sum_{z=0}^{2^m-1}\ket{\x}_n\ket{z}_m.
\end{equation}

\subsubsection{Unitary Operator} Applying the unitary operator $\U_\mathrm{G}(\theta)$ corresponding to a single coefficient $a \in \mathbb{R}$ in the objective function yields
\begin{align}
\frac{1}{\sqrt{2^m|\mathcal{X}|}}\sum_{\x\in\mathcal{X}}\sum_{z=0}^{2^m-1}\ket{\x}_n\ket{z}_m&\\[-2ex]
	&\hspace{-10ex}\xrightarrow{\U_\mathrm{G}(\theta)}\frac{1}{\sqrt{2^m|\mathcal{X}|}}\sum_{\x\in\mathcal{X}}\sum_{z=0}^{2^m-1}\ket{\x}_ne^{\mathrm{j}z\theta}\ket{z}_m,
	\nonumber
\end{align}
where $\mathrm{j} = \sqrt{-1}$ and $\theta = 2\pi a / 2^m \in [-\pi,\pi)$.

The operator $\U_\mathrm{G}(\theta)$ is constructed as a sequence of phase gates, expressed as
\begin{align}
    \U_{\mathrm{G}}(\theta)=\underbrace{\R(2^{m-1}\theta)\otimes\R(2^{m-2}\theta)\otimes\cdots\R(2^0\theta)}_{m~\text{times}},
    \label{eq: UGtheta}
\end{align}
where the phase gate $\R(\theta)$ is defined by
\begin{align}
    \R(\theta)\triangleq\left[
    \begin{array}{cc}
        1 & 0 \\
        0 & e^{\mathrm{j}\theta}
    \end{array}
    \right].
\end{align}

\subsubsection{Inverse Quantum Fourier Transform} Apply the \ac{IQFT} \cite{shor1999polynomialtime} to the lower $m$ qubits, that is
\begin{align}
\frac{1}{\sqrt{2^m|\mathcal{X}|}}\sum_{\x\in\mathcal{X}}\sum_{z=0}^{2^m-1}\ket{\x}_ne^{\mathrm{j}z\theta}\ket{z}_m &\\[-2ex]
&\hspace{-15ex}=\frac{1}{\sqrt{|\mathcal{X}|}}\sum_{\x\in\mathcal{X}}\ket{\x}_n\frac{1}{\sqrt{2^m}}\sum_{z=0}^{2^m-1}e^{\mathrm{j}z\theta}\ket{z}_m\nonumber\\
&\hspace{-13ex}\xrightarrow{\mathrm{IQFT}}\frac{1}{\sqrt{|\mathcal{X}|}}\sum_{\x\in\mathcal{X}}\ket{\x}_n\ket{E(\x)-y}_m.\nonumber
\end{align}

Collecting all the steps outlined above, the state preparation operator, which calculates the value of $E(\x)-y$ in $m$ qubits for any $\x \in \mathbb{B}^n$, can be concisely described as
\begin{align}
\A_y\ket{0}_{n+m}=\frac{1}{\sqrt{|\mathcal{X}|}}\sum_{\x\in\mathcal{X}}\ket{\x}_n\ket{E(\x)-y}_m.
\label{eq:Ay}
\end{align}

Finally, the probability amplitude of the solution state is amplified by applying the gate $\G = \A_y \mathrm{C} \A_y^\dag \mathrm{O}$ for $L_i$ iterations, where $\mathrm{C}$ is the Grover diffusion operator \cite{grover1996fast}, and $\mathrm{O}$ is an oracle that flips the phase of the states satisfying $E(\x)-y < 0$.
Here, the value of $E(\x)-y$ is expressed in two's complement form, such that negative values can be identified by checking the most significant qubit among the $m$ qubits via the Pauli-$\mathrm{Z}$ gate
\begin{align}
\mathrm{Z}=\left[
\begin{array}{cr}
1 & 0 \\
0 & -1
\end{array}
\right].
\end{align}

In the original proposal \cite{gilliam2021grover}, the coefficients of the objective function $E(\x)$ were restricted to integer values.  
It has been shown \cite{norimoto2023quantum}, however, that this limitation can be relaxed to allow real coefficients, at the cost of a slight increase in complexity, by performing classical post-processing after circuit measurement.  
\begin{figure}[H]
\vspace{-2ex}
\centering
\includegraphics[clip, scale=0.47]{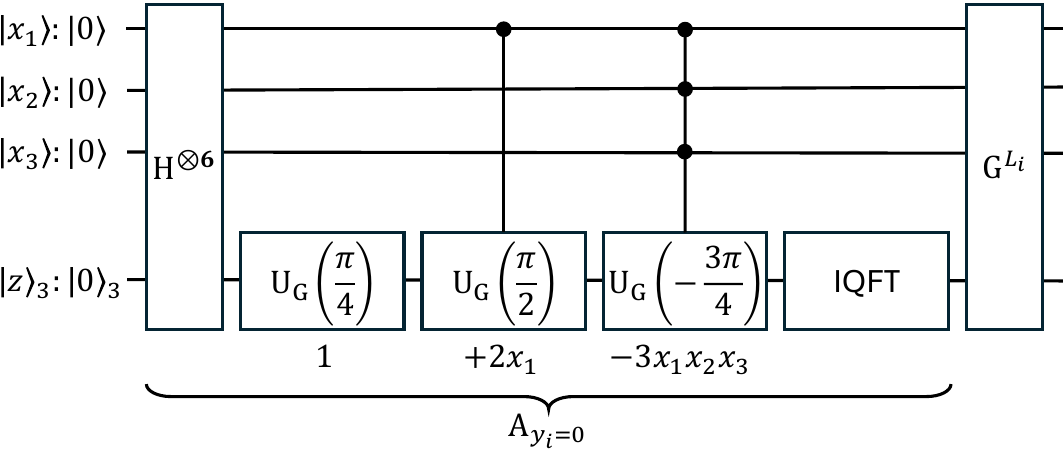}
\vspace{-2ex}
\caption[]{An example of a quantum circuit for the \ac{GAS} algorithm with $E(\x)=1+2x_1-3x_1x_2x_3$ and $y_i=0$.}
\label{fig:GAS}
\end{figure}

As a concrete example, the quantum circuit of the \ac{GAS} algorithm corresponding to the objective function $E(\x)=1+2x_1-3x_1x_2x_3$, with $y_i=0$, is shown in Figure \ref{fig:GAS}.

\section{Proposed QUBO Formulation for QAP}
\label{sec:Proposed QUBO}

In this section, we introduce a method to construct a \ac{PPO} that generates a uniform superposition of all permutation matrices in the reworked \ac{QUBO} formulation. Before doing so, we describe an efficient implementation of the W-state preparation operator \cite{cruz2019efficient} employed in the \ac{PPO}.

\subsection{QUBO Formulation}

By designing the initial state of \ac{GAS} so that the search space already satisfies the constraints in \eqref{BO_obj}, the penalty terms in \eqref{eq:QUBO_obj} become unnecessary, and the objective can be rewritten as
\begin{equation}
    \underset{\mathrm{x}\in\mathcal{P}_n^{\mathrm{Q}}}{\text{minimize}} \quad\underbrace{\left< \mathrm{F}, \mathrm{X} \mathrm{D} \mathrm{X}^{\mathrm{T}} \right>}_{E_\mathrm{Q}(\x)},
    \label{eq:QUBO_obj_permutation}
\end{equation}
where $\mathcal{P}_n^{\mathrm{Q}}$ denotes the set of all permutation patterns corresponding to the \ac{QUBO} formulation (see \ref{subsec: PPO for QUBO}).

\subsection{W State Preparation Operator}

A W state \cite{cruz2019efficient} is defined as the uniform superposition of $n$ basis states, each of which has exactly one qubit in state $1$ and all others in state $0$, and is expressed as
\begin{equation}
    \ket{W_n}=\frac{1}{\sqrt{n}}(\underbrace{\ket{10\cdots0}_n+\ket{01\cdots0}_n+\cdots+\ket{00\cdots1}_n}_{\text{$n$ basis states}}).
\end{equation}

Let $\mathrm{U}_{\mathrm{W}}^n$ denote the unitary operator that generates $\ket{W_n}$ from $n$ qubits initialized to state $0$. The operator $\mathrm{U}_{\mathrm{W}}^n$ performs the following state transition:
\begin{equation}
    \ket{0^n} \xrightarrow{\mathrm{U}_{\mathrm{W}}^n} \ket{W_n}.
\end{equation}

The efficient implementation of $\mathrm{U}_{\mathrm{W}}^n$ has been extensively studied. In particular, a method was proposed in \cite{cruz2019efficient} that exploits a binary tree structure to reduce the circuit depth to approximately $O(\log n)$.

Specifically, the implementation can be achieved by recursively dividing the input state $\ket{10^{n-1}}$ at the root into two branches. For instance, at a given node, the input state $\ket{10^{k-1}}$ consisting of $k$ qubits splits its probability amplitude into a left subtree with $a \triangleq \lceil k/2 \rceil$ qubits and a right subtree with $k-a$ qubits. This branching is expressed by
\begin{equation}
\ket{10^{k-1}}
    \rightarrow \sqrt{\frac{a}{k}} \ket{10^{a-1}}\ket{0^{k-a}} + \sqrt{\frac{k-a}{k}} \ket{0^{a}}\ket{10^{k-a-1}}.
\end{equation}


Here, the first term represents when $\ket{1}$ remains at the head of the left subtree while all right-subtree qubits are $\ket{0}$, whereas the second term occurs when all left-subtree qubits are $\ket{0}$ and $\ket{1}$ appears at the head of the right subtree. This operation is done by applying a $\mathrm{C}R_y(\theta)$ gate followed by a $\mathrm{CNOT}$ to the qubit string, with the first qubit as control and the $(a+1)$-th as target, where $\theta = 2 \arccos(\sqrt{a/k})$.

\begin{figure}[H]
    \vspace{-1ex}
    \centering
    \includegraphics[clip,scale=0.48]{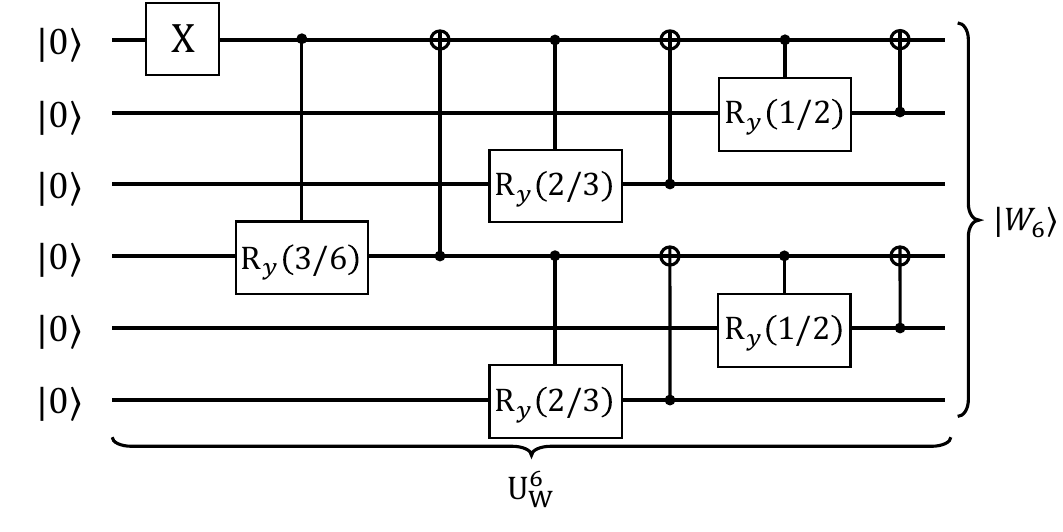}
    \vspace{-3ex}
    \caption{Implementation of $\mathrm{U}_{\mathrm{W}}^6$.}
    \label{fig:U_wstate}
\end{figure}

Figure \ref{fig:U_wstate} illustrates the construction of $\mathrm{U}_{\mathrm{W}}^6$. The fractions shown in $\mathrm{C}\R_y(a/k)$ are used to compute $\theta = 2 \arccos(\sqrt{a/k})$. The state transitions are described below.
First, a Pauli-X gate is applied to generate the state $\ket{100000}$. Next, applying a $\mathrm{C}R_y(3/6)$ gate followed by a $\mathrm{CNOT}$ gate yields
\begin{equation}
    \sqrt{\frac{3}{6}} \ket{100000} + \sqrt{\frac{3}{6}} \ket{000100}.
\end{equation}

Subsequently, applying a $\mathrm{C}R_y(2/3)$ gate and a $\mathrm{CNOT}$ gate yields
\begin{equation}
\begin{split}
\sqrt{\frac{3}{6}\frac{2}{3}} \ket{100000}+\sqrt{\frac{3}{6}\frac{1}{3}} \ket{001000}\\
+ \sqrt{\frac{3}{6}\frac{2}{3}} \ket{000100} + \sqrt{\frac{3}{6}\frac{1}{3}} \ket{000001}.
\end{split}
\end{equation}

Finally, the application of a $\mathrm{C}R_y(1/2)$ gate followed by a $\mathrm{CNOT}$ gate yields the desired state $\ket{W_6}$,
\begin{equation}
\begin{split}
    \sqrt{\frac{3}{6}\frac{2}{3}\frac{1}{2}} \ket{100000} \!+\!\sqrt{\frac{3}{6}\frac{2}{3}\frac{1}{2}} \ket{010000} \!+\! \sqrt{\frac{3}{6}\frac{1}{3}} \ket{001000} \\
    +\sqrt{\frac{3}{6}\frac{2}{3}\frac{1}{2}} \ket{000100} \!+\! \sqrt{\frac{3}{6}\frac{2}{3}\frac{1}{2}} \ket{000010} \!+\! \sqrt{\frac{3}{6}\frac{1}{3}} \ket{000001}.
\end{split}
\end{equation}

By repeating the branching operation at each node, every path from the root to a leaf corresponds to a distinct position of $\ket{1}$. In other words, each leaf node represents a basis state in which $1$ appears at a specific position, and the entire tree forms the superposition state generated by $\mathrm{U}_{\mathrm{W}}^n$. Since the depth of the binary tree is on the order of $\log n$, the circuit depth in this tree-based implementation is likewise reduced to $O(\log n)$.


\subsection{QUBO Permutation Preparation Operator}
\label{subsec: PPO for QUBO}

Let $\mathcal{P}_n$ denote the set of all permutation matrices of size $n$. In \cite{matsuo2023enhancing}, a method was proposed in the context of the \ac{QUBO} formulation to generate a uniform superposition of the states corresponding to all permutation matrices $\X \in \mathcal{P}_n$. This uniform superposition state is defined as 
\begin{equation}
    \ket{P_n^{\mathrm{Q}}}=\frac{1}{\sqrt{n!}}\sum_{\x_\mathrm{Q}\in\mathcal{P}_n^{\mathrm{Q}}} \ket{\x_\mathrm{Q}},
\end{equation}
where the string of qubits $\x_\mathrm{Q}$ is arranged as $\x_\mathrm{Q}=[x_{0,0},x_{0,1},\cdots,x_{0,n-1},\cdots ,x_{n-1,0},\cdots,x_{n-1,n-1}]$, consisting of $n^2$ qubits in a single sequence, and the set $\mathcal{P}_n^{\mathrm{Q}}$ is defined as the collection of bijections obtained by applying the row vectorization $\mathrm{vec}(\X)$ to each permutation matrix $\X \in \mathcal{P}_n$.

For example, when $n=2$, the set of all permutation matrices is
\begin{equation}
\mathcal{P}_2=\Bigg\{\!\begin{bmatrix}1&0\\0&1\end{bmatrix},\begin{bmatrix}0&1\\1&0\end{bmatrix}\!\Bigg\},
\end{equation}
and in this case we have $\x_\mathrm{Q}\in\mathcal{P}_2^\mathrm{Q}=\{[1,0,0,1],[0,1,1,0]\}$.

Define a \ac{PPO} for \ac{QUBO} $\mathrm{U}_{\mathrm{P},\mathrm{Q}}^{n}$ as the unitary operator that generates $\ket{P_{n}^{\mathrm{Q}}}$ on the $n^{2}$ qubits associated with the \ac{QUBO} formulation. The operator can be recursively constructed through the following steps.

\paragraph{State Preparation for $n=2$}

First, we present the construction of $\mathrm{U}_{\mathrm{P},\mathrm{Q}}^{2}$. For the qubit string $[x_{0,0},x_{0,1},x_{1,0},x_{1,1}]$ initialized to $\ket{0}$, $\ket{P_{2}^{\mathrm{Q}}}$ can be generated by performing the following operations.
\begin{equation}
    \begin{split}
        \ket{x_{0,0}x_{0,1}x_{1,0}x_{1,1}}=&\ket{0000}\\
        \xrightarrow{\H(0,0)}&\frac{1}{\sqrt{2}}(\ket{1000}+\ket{0000})\\
        \xrightarrow{\X(0,1)}&\frac{1}{\sqrt{2}}(\ket{1100}+\ket{0100})\\
        \xrightarrow{\mathrm{CNOT}((0,0),(0,1))}&\frac{1}{\sqrt{2}}(\ket{1000}+\ket{0100})\\
        \xrightarrow{\mathrm{CNOT}((0,0),(1,0))}&\frac{1}{\sqrt{2}}(\ket{1001}+\ket{0100})\\
        \xrightarrow{\mathrm{CNOT}((0,1),(1,0))}&\frac{1}{\sqrt{2}}(\ket{1001}+\ket{0110})\\
    \end{split}
\end{equation}

\begin{figure}[t]
    \centering
    \includegraphics[clip,scale=0.48]{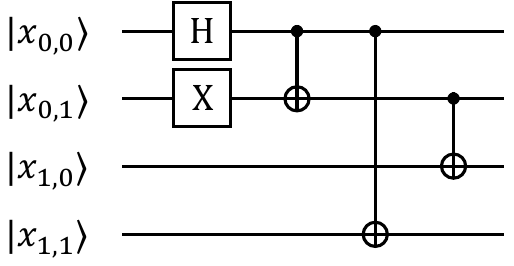}
    \caption{Construction of $\mathrm{U}_\mathrm{P,Q}^2$. }
    \label{fig:QAP_N=2}
\end{figure}

Figure \ref{fig:QAP_N=2} shows the quantum circuit that implements $\mathrm{U}_{\mathrm{P},\mathrm{Q}}^{2}$.

\paragraph{State Preparation for $n=k$ $(k\geq 3)$}

Next, we consider the extension from $\ket{P_{k-1}^{\mathrm{Q}}}$ to $\ket{P_{k}^{\mathrm{Q}}}$. Assume that $\ket{P_{k-1}^{\mathrm{Q}}}$ has already been prepared by applying $\mathrm{U}_{\mathrm{P},\mathrm{Q}}^{k-1}$ to the $(k-1)^{2}$ qubits $[x_{0,0},\cdots,x_{0,k-2},x_{1,0}\cdots,x_{k-2,k-2}]$, and then apply the operator $\mathrm{U}_{\mathrm{W}}^{k}$ to the qubits $[x_{k-1,0},\cdots,x_{k-1,k-1}]$ corresponding to the $k$-th row of a size-$k$ permutation matrix to prepare the W state $\ket{W_k}$ \cite{cruz2019efficient}.

\begin{figure}[t]
\vspace{-4ex}
\centering
\subfigure[Step $1$]{
\includegraphics[clip, width=\columnwidth]{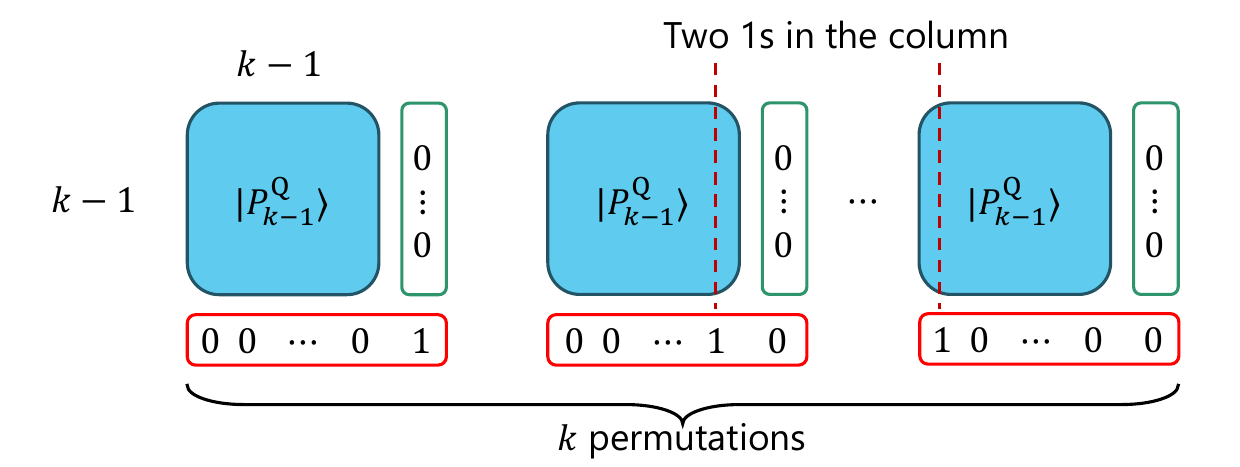}}
\subfigure[Step $2$]{
\includegraphics[clip, width=\columnwidth]{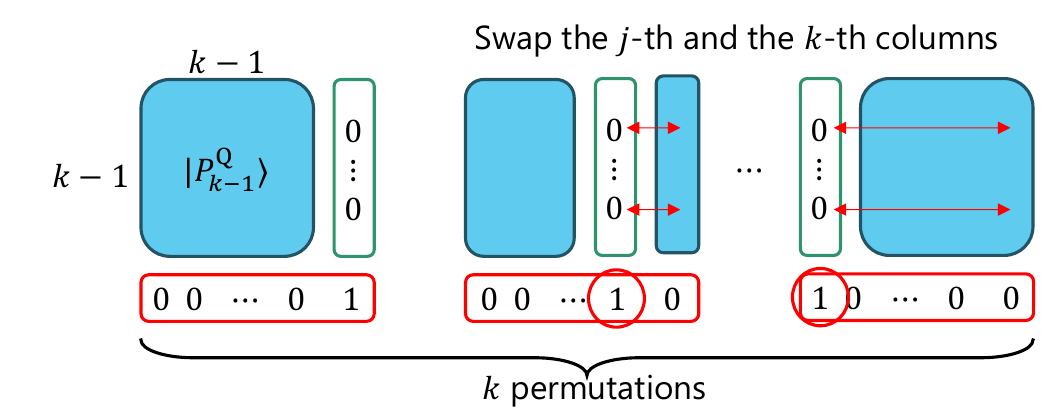}}
\caption{Expansion from $\ket{P_{k-1}^{\mathrm{Q}}}$ to $\ket{P_{k}^{\mathrm{Q}}}$.}
\label{fig:PPO_QUBO_steps}
\end{figure}

This operation is crucial for generating the set of $k \times k$ permutation matrices, since the $k$-th row itself represents a permutation. At this stage, as shown in Figure \ref{fig:PPO_QUBO_steps} (a), $\ket{P_{k-1}^\mathrm{Q}}$ is divided into $k$ branches depending on the position of the $1$ in the $k$-th row.

Here, the bit string $\x_\mathrm{Q}$ belongs to $\mathcal{P}_k^\mathrm{Q}$ if and only if $x_{k-1,k-1}=1$. Otherwise, \textit{i.e.} when $x_{k-1,j}=1$ with $0 \leq j \leq k-2$, the constraint that each row and each column of $\X$ must contain exactly one entry equal to $1$ is violated, and thus $\x_\mathrm{Q}$ no longer belongs to $\mathcal{P}_k^\mathrm{Q}$.

To resolve this issue, perform the operation shown in Figure~\ref{fig:PPO_QUBO_steps}(b): when $x_{k-1,j}=1$ for $0 \le j \le k-2$, swap $x_{i,j}$ and $x_{i,k-1}$ for all $i$ such that $0 \le i \le k-2$. This operation is implemented using controlled-$\mathrm{SWAP}$ ($\mathrm{CSWAP}$) gates, where $x_{k-1,j}$ serves as the control bit and $x_{i,j},x_{i,k-1}$ serve as the target bits. The collection of these $(k-1)^2$ operations is denoted by $\mathrm{U}_{\mathrm{CSWAP}}^{k}$. As a result, for any $j$ with $0 \leq j \leq k-1$, when $x_{k-1,j}=1$, the state $\ket{P_{k-1}^{\mathrm{Q}}}$ is generated on the $(k-1)^2$ qubits excluding the $k$-th row and $j$-th column, so that the overall state $\ket{P_{k}^{\mathrm{Q}}}$ is obtained (see Figure~\ref{fig:PPO_QUBO_steps}).

In summary, to construct the permutation preparation operator $\mathrm{U}_{\mathrm{P},\mathrm{Q}}^{N}$ for problem size $N$, first apply $\mathrm{U}_{\mathrm{P},\mathrm{Q}}^{2}$ to $[x_{0,0},x_{0,1},x_{1,0},x_{1,1}]$ to generate $\ket{P_{2}^{\mathrm{Q}}}$. Then, for $k=3,\dots,N$, the following steps are repeated.
\begin{itemize}
\item $\mathrm{U}_{\mathrm{W}}^{k}$ is applied to the $k$ qubits $[x_{k-1,0},\dots,x_{k-1,k-1}]$,
\item $\mathrm{U}_{\mathrm{CSWAP}}^{k}$ is applied to the $k^{2}$ qubits $x_{i,j}~(0 \leq i,j \leq k-1)$.
\end{itemize}

\begin{figure}[H]
    \centering
    \includegraphics[clip,scale=0.48]{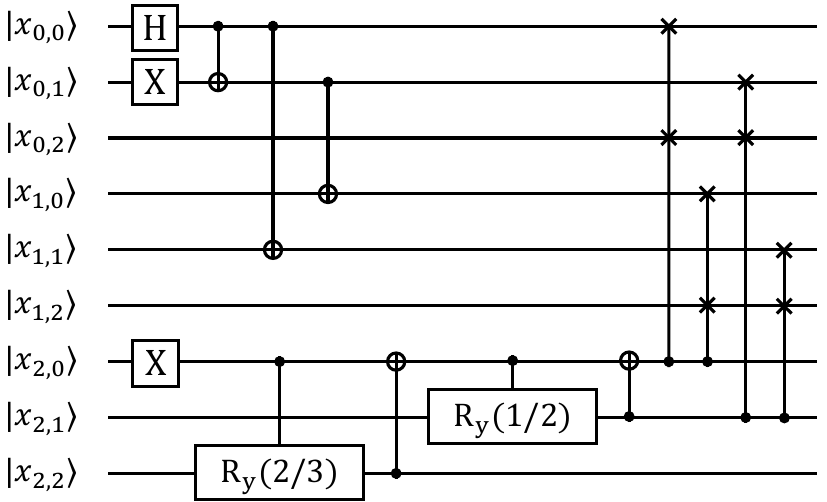}
\caption{Construction of $\mathrm{U}_\mathrm{P,Q}^3$.\label{fig:PPO_N=3}}
\end{figure}

For reference, Figure \ref{fig:PPO_N=3} illustrates the construction of $\mathrm{U}_{\mathrm{P},\mathrm{Q}}^{3}$ in the case $N=3$.

\subsubsection{Integrating PPO for QUBO into GAS}

In order to integrate \ac{PPO} for \ac{QUBO} into \ac{GAS}, the uniform superposition state prepared by the state preparation operator $\A_y$ must be replaced. Thus, the circuit construction method in Sec. \ref{subsubsec: initial state preparation of GAS} is revised as follows.

%

Start by applying \ac{PPO} for \ac{QUBO} $\mathrm{U}_{\mathrm{P},\mathrm{Q}}^{N}$ to $n=N^2$ qubits so that the initial state $\ket{0}_{n+m}$ is transformed into $\ket{P_{N}^{\mathrm{Q}}}$, with the corresponding transition described by
\begin{equation}
\ket{0}_{n+m}\xrightarrow{\mathrm{U}_{\mathrm{P},\mathrm{Q}}^{N}} \ket{P_{N}^{\mathrm{Q}}}\ket{0}_m =\frac{1}{\sqrt{N!}}\sum_{\x_{\mathrm{Q}}\in\mathcal{P}_N^\mathrm{Q}}\ket{\x_\mathrm{Q}}_n\ket{0}_m.
\end{equation}

Then, as with the original \ac{GAS}, Hadamard gates are applied to the least significant $m$ qubits, yielding the transition
\begin{equation}
\begin{split}
&\frac{1}{\sqrt{N!}}\sum_{\x_{\mathrm{Q}}\in\mathcal{P}_N^\mathrm{Q}}\ket{\x_\mathrm{Q}}_n\ket{0}_m\\
&\xrightarrow{\H^{\otimes m}} \frac{1}{\sqrt{2^m N!}}\sum_{\x_{\mathrm{Q}}\in\mathcal{P}_N^\mathrm{Q}}\sum_{z=0}^{2^m-1}\ket{\x_\mathrm{Q}}_n\ket{z}_m.
\end{split}
\end{equation}

The remaining steps are the same as those presented in Sec. \ref{subsec:GAS}. As a result, the transition produced by the state preparation operator $\A_y$ is modified from \eqref{eq:Ay} to 
\begin{align}
\A_y\ket{0}_{n+m}=\frac{1}{\sqrt{N!}}\sum_{\x_\mathrm{Q}\in\mathcal{P}_N^\mathrm{Q}}\ket{\x_\mathrm{Q}}_n\ket{E_\mathrm{Q}(\x_\mathrm{Q})-y}_m.
\end{align}

\section{Proposed HUBO Formulation for QAP}
\label{sec:Proposed HUBO}

In this section, we propose a formulation of the \ac{QAP} as a \ac{HUBO} problem, together with a corresponding \ac{PPO}, with the aim of solving the \ac{QAP} with fewer qubits.

\subsection{HUBO Formulation}

In the conventional \ac{QUBO} formulation, $N^{2}$ binary variables are required, which is inefficient compared to the lower bound of $O(N \log N)$, as indicated in \eqref{eq:log2Nf}. To address this issue, we propose a higher-order formulation that reduces the number of binary variables to $N \lceil \log_{2} N \rceil$. However, this approach has the drawback that the degree of the objective function exceeds quadratic.

Regarding the solution matrix $\mathrm{X} \in \mathbb{B}^{N \times N}$, it can be viewed as a set of $N$ indices, each of which can be represented as a one-hot vector. By encoding these indices in binary form, the number of variables can be reduced.
In related studies \cite{glos2022space, tabi2020quantum}, \ac{QUBO} problems employing one-hot encoding have been reformulated as \ac{HUBO} problems by mapping the indices to binary representations.

In the \ac{QAP}, indices are assigned to both facilities and locations. Here, we focus on encoding the location indices in binary. For a given problem size $N$, the binary vector corresponding to the location index $j$ ($0 \leq j \leq N-1$) is defined as
\begin{equation}
    \b_j=[b_{j,0}, b_{j,1}, \cdots , b_{j,B-1}]=[j]_2,
\end{equation}
where the length of the vector is
\begin{equation}
    B\triangleq \lceil\log_2 N\rceil
\end{equation}
and $[\cdot]_2$ denotes the conversion from decimal to binary. 

Since little-endian representation is used, the following equality holds.
\begin{equation}
    j=\sum_{k=0}^{B-1} b_{j,k}\cdot 2^{k}
\end{equation}

By using this representation, the solution matrix $\Y \in \mathbb{B}^{N \times N}$ in the \ac{HUBO} formulation of \ac{QAP} is encoded by a new set of $NB$ binary variables $x_{ir}$ $(0 \leq i \leq N-1,~ 0 \leq r \leq B-1)$. The $(i,j)$-th entry of $\Y$ is then given by
\begin{equation}
    \label{eq:delta_ij}
    y_{i,j}(x) = \prod_{r=0}^{B-1}(1-b_{j,r}+(2b_{j,r}-1)x_{i,r})
\end{equation}

\begin{table}[t]
\centering
\caption{Example of binary encoding.}
\begin{tabular}{lcl}
\hline
$j$ & $[b_{j,0} ~ b_{j,1}]$ & $y_{ij}(x)$\\
\hline
$0$ & $[0~0]$& $(1-x_{i,0})(1-x_{i,1})$\\
$1$ & $[1~0]$& $x_{i,0}(1-x_{i,1})$\\
$2$ & $[0~1]$& $(1-x_{i,0})x_{i,1}$\\
$3$ & $[1~1]$& $x_{i,0}x_{i,1}$\\
\hline
\end{tabular}
\label{tab:HUBO_ASC}
\end{table}

As an example, the case $N=4$ is shown in Table \ref{tab:HUBO_ASC}. When the $i$-th facility is assigned to location $j=2$, that is, when $[x_{i,0}, x_{i,1}] = [0,1]$ and only in this case, the corresponding entry is given by $y_{i,2}(x) = (1-x_{i,0})x_{i,1} = 1$, while for the other assignments $[x_{i,0}, x_{i,1}] \in \{[0,0],[1,0],[1,1]\}$, we have $y_{i,2}(x) = 0$.

For reference, the solution matrix for $N=3$ is given by
\begin{equation}
    \Y\!=\!\left[\!\!
    \begin{array}{ccc}
    (1-x_{0,0})(1-x_{0,1}) \!\!\!& x_{0,0}(1-x_{0,1}) \!\!\!& (1-x_{0,0})x_{0,1}\\
    (1-x_{1,0})(1-x_{1,1}) \!\!\!& x_{1,0}(1-x_{1,1}) \!\!\!& (1-x_{1,0})x_{1,1}\\
    (1-x_{2,0})(1-x_{2,1}) \!\!\!& x_{2,0}(1-x_{2,1}) \!\!\!& (1-x_{2,0})x_{2,1}
    \end{array}
    \!\!\right]
    \label{eq:matrixY}
\end{equation}

Finally, the objective function of \ac{HUBO} formulation is given by 
\begin{equation}
    \underset{\x\in\mathbb{B}^{N\times B}}{\text{minimize}}\quad \underbrace{\left< \mathrm{F}, \mathrm{Y} \mathrm{D} \mathrm{Y}^{\mathrm{T}} \right>}_{E_\mathrm{H}(\x)}.
    \label{eq:HUBO_obj_permutation}
\end{equation}

Note that the constraint requiring exactly one entry of $1$ in each row and each column of $\Y$ is not included in the objective as a penalty term. This constraint is enforced by the \ac{PPO} corresponding to the \ac{HUBO} formulation presented in \ref{subsec: PPO for HUBO}.

\begin{figure}[h]
    \centering
    \includegraphics[clip, scale=0.48]{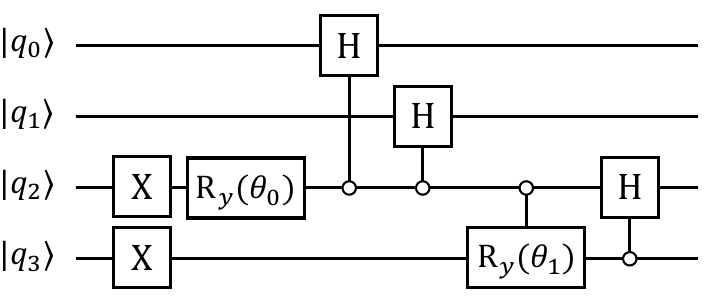}
\caption{Construction of $\U_{\mathrm{S}}^{13,4}$.\label{fig:Shukla}}
\end{figure}

\begin{algorithm}[H]
\caption{Construction of Shukla state prep. operator \cite{shukla2024efficient}.\label{alg:shukla}}
\begin{algorithmic}[1]
\renewcommand{\algorithmicrequire}{\textbf{Input:}}
\renewcommand{\algorithmicensure}{\textbf{Output:}}
\Require {A positive integer $M$ with $2<M<2^n$ and $M\neq 2^r$ for any $r\in\mathbb{N}$}
\Ensure $\ket{\U_\mathrm{S}^{n,M}}$
\State Compute $\ell_0,\cdots \ell_k$, where $M=\sum_{j=0}^k 2^{\ell_i}$ with 
\Statex\hspace{\algorithmicindent} $0\leq \ell_0< \ell_1<\cdots<\ell_k\leq n-1$.
\State Initialzie $\ket{q_{0}}\otimes\ket{q_{1}}\otimes\cdots\otimes\ket{q_{n-1}}=\ket{0}^{\otimes n}$.
\State Apply $\X$ gate on $\ket{q_i}$ for $i=\ell_1,\ell_2,\cdots,\ell_k$.
\State $M_0=2^{\ell_0}$.
\hspace{\algorithmicindent}\If{$\ell_0>0$}
\State Apply $\H$ gate on $\ket{q_i}$ for $i=0,1,\cdots,\ell_0-1$
\EndIf
\State Apply rotation gate $\R_y(\theta_0)$ on $\ket{q_{\ell_1}}$, where 
\Statex \hspace{\algorithmicindent} $\theta_0=-2\arccos(\sqrt{\frac{M_0}{M}})$.
\State Apply a $\mathrm{C}\H$ gate on $\ket{q_i}$ for $i=\ell_0,\ell_0+1,\cdots,\ell_1-1$
\Statex \hspace{\algorithmicindent} conditioned on $q_{\ell_1}$ being equal to $0$.
\For{$m=1$ to $k-1$}
\State Apply a $\mathrm{C}\R_y(\theta_m)$ gate on $\ket{q_{\ell_{m+1}}}$ controlled by 
\Statex \hspace{\algorithmicindent} $\ket{q_{\ell_m}}$ being $0$ with $\theta_m=-2\arccos(\sqrt{\frac{2^{\ell_m}}{M-M_{m-1}}})$
\State {\small{Apply a $\mathrm{C}\H$ gate on $\ket{q_i}$ for $i=\ell_m,\ell_m+1,\cdots,\ell_{m+1}-1$}}
\Statex \hspace{\algorithmicindent}conditioned on $\ket{q_{\ell_{m+1}}}$ being $\ket{0}$
\State Set $M_m=M_{m-1}+2^{\ell_m}$
\EndFor
\end{algorithmic}
\end{algorithm}

In the next subsection, we describe the construction method of the Shukla state preparation operator used in the \ac{PPO} corresponding to the \ac{HUBO} formulation.

\subsection{Shukla State Preparation Operator}

A Shukla state \cite{shukla2024efficient}, proposed by Shukla et al., is a uniform superposition state over $n$ qubits for any integer $M$ with $M < 2^n$, defined as 
\begin{equation}
    \ket{S_M}=\frac{1}{\sqrt{M}}\sum_{i=0}^{M-1} \ket{i}_n.
\end{equation}

Algorithm \ref{alg:shukla} presents the construction of the Shukla state preparation operator $\U_\mathrm{S}^{M,n}$, a unitary operator that generates the Shukla state corresponding to an integer $M$ on $n$ qubits. In the special case $M=2^r$ ($\forall r \in \mathbb{N}$), it suffices to apply Hadamard gates to the $r$ qubits $\ket{q_0}, \cdots, \ket{q_{r-1}}$. The procedure of the algorithm is described below.

First, prepare $n$ qubits $(\ket{q_0}, \cdots, \ket{q_{n-1}})$, each initialized in the state $\ket{0}$.
Define $M$ as a natural number satisfying $2 < M < 2^n$ and $M \neq 2^r \; (\forall r \in \mathbb{N})$.
$M$ can be expressed as
\begin{equation}
M=\sum_{i=0}^k 2^{\ell_i},
\end{equation}
where the bit string $[\ell_0,\ell_1,\cdots,\ell_k]$ corresponds to the positions of the ones in the binary representation of $M$, listed in reverse order. For example, when $M=13$, we have $[13]_2=[1,1,0,1]$, and reversing this gives $[1,0,1,1]$. Hence, the corresponding bit string is $[\ell_0,\ell_1,\ell_2]=[0,2,3]$. Using the bit string obtained in this way, apply Pauli-X gates to the qubits $\ket{q_i}$ with $i=\ell_1,\cdots,\ell_k$.
When $M$ is even, that is, when $\ell_0 > 0$, apply Hadamard gates to the $\ell_0$ qubits $\ket{q_0}, \cdots, \ket{q_{\ell_0-1}}$.

Here, we define $M_0 = 2^{\ell_0}$ and apply a rotation gate $\R_y(\theta_0)$ to $\ket{q_{\ell_0}}$, where 
\begin{equation}
\theta_0 = -2 \arccos(M_0/M).
\end{equation}

Next, apply zero-controlled Hadamard gates to the qubits $\ket{q_{\ell_i}}$ $(i=\ell_0, \ell_0+1, \cdots, \ell_1-1)$, using $\ket{q_{\ell_1}}$ as the control qubit.
Then, for $m=1,\cdots,k-1$, repeat the following procedure.
Apply a zero-controlled rotation gate $\mathrm{C}\R_y(\theta_m)$ to $\ket{q_{\ell_{m+1}}}$ with $\ket{q_{\ell_m}}$ as the control qubit, where
\begin{equation}
\theta_m = -2\arccos\!\left(\sqrt{\tfrac{2^{\ell_m}}{M - M_{m-1}}}\right).
\end{equation}

In addition, apply zero-controlled Hadamard gates to the qubits $\ket{q_i}$ with $i=\ell_m, \ell_m+1, \cdots, \ell_{m+1}-1$, using $\ket{q_{\ell_{m+1}}}$ as the control qubit. Finally, update the parameter as $M_m = M_{m-1} + 2^{\ell_m}$.

For the case $n=4$ and $M=13$, the corresponding quantum circuit is shown in Figure \ref{fig:Shukla}. First, following line 3 of Algorithm \ref{alg:shukla}, apply Pauli-X gates to $\ket{q_{\ell_1}}=\ket{q_2}$ and $\ket{q_{\ell_2}}=\ket{q_3}$. The operation in line 6 is omitted because $\ell_0=0$. Next, apply $\R_y(\theta_0)$ to $\ket{q_{\ell_1}}=\ket{q_2}$, where $M_0=2^{\ell_0}=2^0=1$ and $\theta_0=-2\arccos\big(\sqrt{1/13}\big)$. Furthermore, apply zero-controlled Hadamard gates to $\ket{q_i}$ for $i=\ell_0,\cdots,\ell_1-1=0,1$, using $\ket{q_{\ell_1}}=\ket{q_2}$ as the control qubit.

We enter the loop starting at line 10. Since $k=2$ in this case, the loop is executed only once. Apply a zero-controlled rotation gate $\mathrm{C}\R_y(\theta_1)$ to $\ket{q_{\ell_2}}=\ket{q_3}$ with $\ket{q_{\ell_1}}=\ket{q_2}$ as the control qubit, where
\begin{equation}
\theta_1=-2\arccos\!\left(\sqrt{\tfrac{2^{\ell_1}}{M-M_{0}}}\right)=-2\arccos\!\left(\sqrt{\tfrac{4}{12}}\right).
\end{equation}

Finally, apply zero-controlled Hadamard gates to $\ket{q_i}$ for $i=\ell_1,\cdots,\ell_{m+1}-1=2$, using $\ket{q_{\ell_2}}=\ket{q_3}$ as the control qubit. In this way, the desired operator $\U_{\mathrm{S}}^{13,4}$ is successfully constructed.


\subsection{HUBO Permutation Preparation Operator}
\label{subsec: PPO for HUBO}

The uniform superposition state over all permutation patterns of size $n$ corresponding to the \ac{HUBO} formulation is defined as 
\begin{equation}
    \ket{P_n^{\mathrm{H}}}=\frac{1}{\sqrt{n!}}\sum_{\x_\mathrm{H}\in\mathcal{P}_n^{\mathrm{H}}} \ket{\x_\mathrm{H}},
\end{equation}
where the qubit string $\x_\H$ consists of $nB$ qubits, $\x_\mathrm{H} = [x_{0,0}, \cdots, x_{0,B-1}, \cdots, x_{n-1,0}, \cdots, x_{n-1,B-1}]$ , arranged in a single sequence, assuming that $n\leq 2^B$.

\begin{algorithm}[H]
\caption{Construction of permutation preparation operator for HUBO.\label{alg:PPO for HUBO}}
\begin{algorithmic}[1]
\renewcommand{\algorithmicrequire}{\textbf{Input:}}
\renewcommand{\algorithmicensure}{\textbf{Output:}}

\Require A positive integer $N$ with $2\leq N$
\Ensure $\mathrm{U}_{\mathrm{P},\mathrm{H}}^{n}$
\State Apply a $\H$ gate to $\ket{x_{0,0}}$.
\State Apply a $\X$ gate to $\ket{x_{1,0}}$.
\State Apply a $\mathrm{CNOT}$ gate to $\ket{x_{1,0}}$ with $\ket{x_{0,0}}$ as control.
\For{$k=3$ to $n$}
\State Apply a Shukla state preparation operator $\mathrm{U}_\mathrm{S}^{k,B}$ to the $B$ qubits $[x_{k-1,0}, \cdots, x_{k-1,B-1}]$.
\For{$\ell=0$ to $k-2$}
\For{$i=0$ to $k-2$}
\State apply a $\mathrm{C}^{2B}\X$ gate to $\ket{\text{ancilla}}$ with
\Statex\hspace{\algorithmicindent}\hspace{\algorithmicindent} ~~~$[x_{i,0},\cdots,x_{i,B-1}]$ and $[x_{k-1,0},\cdots,x_{k-1,B-1}]$
\Statex\hspace{\algorithmicindent}\hspace{\algorithmicindent} ~~~as control qubits \plainfootmark
\State Apply a $\mathrm{CNOT}$ gate to $\ket{x_{i,j}}$ with $\ket{\text{ancilla}}$ 
\Statex\hspace{\algorithmicindent}\hspace{\algorithmicindent} ~~~as a control qubit, 
\Statex\hspace{\algorithmicindent}\hspace{\algorithmicindent} ~~~$\forall j \in \{\,0,\dots,B-1 \mid \mathrm{bin}_B(\ell)_j \neq \mathrm{bin}_B(k)_j \,\}$.
\State apply a $\mathrm{C}^{2B}\X$ gate to $\ket{\text{ancilla}}$ with
\Statex\hspace{\algorithmicindent}\hspace{\algorithmicindent} ~~~$[x_{i,0},\cdots,x_{i,B-1}]$ and $[x_{k-1,0},\cdots,x_{k-1,B-1}]$
\Statex\hspace{\algorithmicindent}\hspace{\algorithmicindent} ~~~as control qubits \plainfootmark
\EndFor
\EndFor
\EndFor
\end{algorithmic}
\end{algorithm}

\begin{figure}[H]
    \centering
    \includegraphics[clip,scale=0.48]{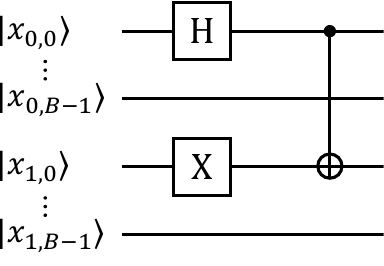}
    \caption{Construction of $\mathrm{U}_\mathrm{P,H}^2$.\label{fig:QAP_HUBO_N=2}}
\end{figure}

\addtocounter{footnote}{-1}
\plainfoottext{In Qiskit, big-endian bit strings are used as control state, where a bit value of 0 specifies a zero-control and a bit value of 1 specifies a one-control. In this case, the control state is given by $\text{reverse}(\text{concat}(\mathrm{bin}_B(\ell),\mathrm{bin}_B(\ell)))$.}
\stepcounter{footnote}
\plainfoottext{The control state is given by $\text{reverse}(\text{concat}(\mathrm{bin}_B(k),\mathrm{bin}_B(\ell)))$ in this case.}

To define $\mathcal{P}_n^{\mathrm{H}}$, we first introduce several definitions. Let $\mathfrak{S}_n$ denote the symmetric group, \textit{i.e.} the set of all permutations of the set $\{0,1,\cdots,n-1\}$. For example, when $n=3$,
\begin{equation}
    \mathfrak{S}_3\!=\!\!\{(0,\!1,2),\!(0,2,\!1),\!(1,0,2),\!(1,2,0),\!(2,0,\!1),\!(2,\!1,0)\!\}
\end{equation}

For a permutation pattern $(s_0, s_1, \cdots, s_{n-1}) \in \mathfrak{S}_n$, define the operator $\mathrm{bin}_B(s_i) = [b_{i,0}, b_{i,1}, \cdots, b_{i,B-1}]$ as the $B$-bit binary conversion of each element $s_i$ in the set. Here, $\mathrm{bin}_B(s_i)$ is expressed in the little-endian form and satisfies
\begin{equation}
    s_i = \sum_{j=0}^{B-1} b_{i,j} \cdot 2^{j}.
\end{equation}

For example, when $B=4$ and $s_i=4$, we have $\mathrm{bin}_B(s_i) = [0,0,1,0]$.

Using the above, $\mathcal{P}_n^{\mathrm{H}}$ is defined as
\begin{equation}
    \mathcal{P}_n^{\mathrm{H}}=\{[\mathrm{bin}_B(s_0),\cdots,\mathrm{bin}_B(s_{n-1})]|\{s_0,\cdots,s_{n-1}\}\in\mathfrak{S}_n\}
\end{equation}

For example, when $n=3$ and $B=2$, $\mathcal{P}_3^{\mathrm{H}}$ is given by
\begin{align}
\mathcal{P}_3^{\mathrm{H}} =&\{[\underbrace{0,0}_0,\underbrace{1,0}_1,\underbrace{0,1}_2],[\underbrace{0,0}_0,\underbrace{0,1}_2,\underbrace{1,0}_1],\nonumber\\
    &~~[\underbrace{1,0}_1,\underbrace{0,0}_0,\underbrace{0,1}_2],[\underbrace{1,0}_1,\underbrace{0,1}_2,\underbrace{0,0}_0],\nonumber\\
    &~~[\underbrace{0,1}_2,\underbrace{0,0}_0,\underbrace{1,0}_1],[\underbrace{0,1}_2,\underbrace{1,0}_1,\underbrace{0,0}_0]
    \}
\label{eq:P_3^H}
\end{align}

Since the size of the original symmetric group $\mathfrak{S}_n$ is $n!$, and each element of $\mathfrak{S}_n$ is in one-to-one correspondence with an element of $\mathcal{P}_n^{\mathrm{H}}$, it follows that $|\mathcal{P}_n^{\mathrm{H}}| = n!$.

We define the \ac{PPO} for \ac{HUBO}, denoted as $\mathrm{U}_{\mathrm{P},\mathrm{H}}^{n}$, as the unitary operator that generates $\ket{P_{n}^{\mathrm{H}}}$ on $nB$ qubits corresponding to the HUBO formulation. The construction method of the operator $\mathrm{U}_{\mathrm{P},\mathrm{H}}^{n}$ is shown in Algorithm~\ref{alg:PPO for HUBO}. We introduce the recursive construction below.

\paragraph{State Preparation for $n=2$}

First, we present the construction of $\mathrm{U}_{\mathrm{P},\mathrm{H}}^{2}$. For the $2B$ qubits string $[x_{0,0},\cdots,x_{0,B-1},\,x_{1,0},\cdots,x_{1,B-1}]$ initialized to $0$, $\ket{P_{2}^{\mathrm{H}}}$ can be generated by 

\quad\\[-6ex]
\begin{eqnarray}
 \ket{x_{0,0}\cdots x_{0,B-1}} \ket{x_{1,0} \cdots x_{1,B-1}} = \ket{0^{B}}\ket{0^B}&&\\
&&\hspace{-38ex}\xrightarrow{\H(0,0)}\frac{1}{\sqrt{2}}(\ket{10^{B-1}}\ket{0^{B}}+\ket{0^{B}}\ket{0^B})\nonumber\\
&&\hspace{-38ex}\xrightarrow{\X(1,0)}\frac{1}{\sqrt{2}}(\ket{10^{B-1}}\ket{10^{B-1}}+\ket{0^{B}}\ket{10^{B-1}})\nonumber\\
&&\hspace{-38ex}\xrightarrow{\mathrm{CNOT}((0,0)(1,0))}\frac{1}{\sqrt{2}}(\ket{\underbrace{10^{B-1}}_{1}}\ket{\underbrace{0^{B}}_0}+\nonumber\\
&&\hspace{-10ex}\ket{\underbrace{0^{B}}_0}\ket{\underbrace{10^{B-1}}_1}).\nonumber
\end{eqnarray}

Figure \ref{fig:QAP_N=2} shows the quantum circuit that implements $\mathrm{U}_{\mathrm{P},\mathrm{H}}^{2}$.

\paragraph{State Preparation for $n=k$ $(k\geq 3)$}

Next, we consider the extension from $\ket{P_{k-1}^{\mathrm{H}}}$ to $\ket{P_{k}^{\mathrm{H}}}$. Assume that $\ket{P_{k-1}^{\mathrm{H}}}$ has already been prepared by applying $\mathrm{U}_{\mathrm{P},\mathrm{H}}^{k-1}$ to the $(k-1)B$ qubits $[x_{0,0},\cdots,x_{0,B-1},\,x_{1,0},\cdots,x_{k-2,B-1}]$.

Next, apply the operator $\mathrm{U}_\mathrm{S}^{k,B}$ to the $B$ qubits $[x_{k-1,0}, \cdots, x_{k-1,B-1}]$ corresponding to row $k-1$ to prepare the Shukla state $\ket{S_k}$ \cite{shukla2024efficient}.

Here, if $[x_{k-1,0},\cdots,x_{k-1,B-1}]=\mathrm{bin}_B(k-1)$, then the bit string $\x_\mathrm{H}$ belongs to the set $\mathcal{P}_{k}^{\mathrm{H}}$. Otherwise, if $[x_{k-1,0},\cdots,x_{k-1,B-1}]=\mathrm{bin}_B(\ell)$ with $0\leq \ell\leq k-2$, duplicates occur in cases where $[x_{i,0},\cdots,x_{i,B-1}]=\mathrm{bin}_B(\ell)$ for some $0\le i\le k-2$. For example, considering the extension from $n=2$ to $n=3$ with $B=2$, the starting set is given by
\begin{equation}
    \mathcal{P}_2^{\mathrm{H}}
    = \big\{[\underbrace{0,0}_{0},\,\underbrace{1,0}_{1}],\;[\underbrace{1,0}_{1},\,\underbrace{0,0}_{0}]\big\},
\end{equation}
and by applying $\mathrm{U}_\mathrm{S}^{3,2}$ to $[x_{2,0},x_{2,1}]$, we obtain the following six basis states.
\begin{equation}
\begin{split}
    &\{[\underbrace{0,0}_0,\underbrace{1,0}_1,\underbrace{0,0}_0],[\underbrace{0,0}_0,\underbrace{1,0}_1,\underbrace{1,0}_1],[\underbrace{0,0}_0,\underbrace{1,0}_1,\underbrace{0,1}_2],\\
    &~~[\underbrace{1,0}_1,\underbrace{0,0}_0,\underbrace{0,0}_0],[\underbrace{1,0}_1,\underbrace{0,0}_0,\underbrace{1,0}_1],[\underbrace{1,0}_1,\underbrace{0,0}_0,\underbrace{0,1}_2]\}.
\end{split}
\label{eq:P_n^H before}
\end{equation}

\begin{figure}[h]
\centering
\includegraphics[clip,scale=0.48]{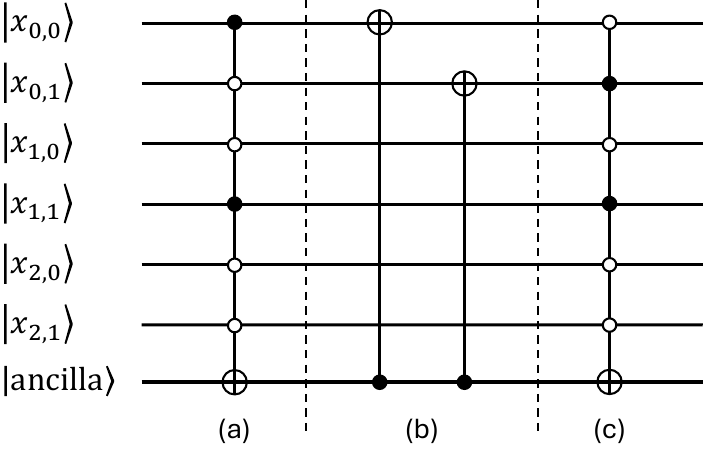}
    \caption{An example of rewriting to the basis states of $\mathcal{P}_k^{\mathrm{H}}$}
    \label{fig: basis transition}
\vspace{-3ex}
\end{figure}

\noindent
In this case, $[\underbrace{0,0}_{0},\,\underbrace{1,0}_{1},\,\underbrace{0,1}_{2}]$ and $[\underbrace{1,0}_{1},\,\underbrace{0,0}_{0},\,\underbrace{0,1}_{2}]$ belong to $\mathcal{P}_3^{\mathrm{H}}$, whereas the remaining four patterns contain duplicate entries (either $0$ or $1$).

For the duplicate patterns, whenever both $[x_{i,0},\cdots,x_{i,B-1}]$ and $[x_{k-1,0},\cdots,x_{k-1,B-1}]$ are equal to $\mathrm{bin}_B(\ell)$, we transform $[x_{i,0},\cdots,x_{i,B-1}]$ into $\mathrm{bin}_B(k-1)$. This ensures that every basis state belongs to $\mathcal{P}_k^{\mathrm{H}}$. The construction of a quantum circuit that implements this operation is described below.

First, prepare an ancillary qubit $\ket{\text{ancilla}}$ to record whether $[x_{i,0},\cdots,x_{i,B-1}]$ and $[x_{k-1,0},\cdots,x_{k-1,B-1}]$ are both equal to $\mathrm{bin}_B(\ell)$. Then, apply a multi-controlled $\mathrm{X}$ gate to $\ket{\text{ancilla}}$, using $[x_{i,0},\cdots,x_{i,B-1}]$ and $[x_{k-1,0},\cdots,x_{k-1,B-1}]$ as control bits. Here, for the $j$-th element of $\mathrm{bin}_B(\ell)$ ($0 \leq j \leq B-1$), if the element is $0$, the corresponding $x_{i,j}$ and $x_{k-1,j}$ are used as zero-controls, otherwise, if the element is $1$, they are used as one-controls.

For example, when $B=3$, $k=3$, and $\ell=1$, this multi-controlled $\mathrm{X}$ gate takes the form shown in Figure \ref{fig: basis transition} (a).

Next, for each $j$ such that the $j$-th elements of $\mathrm{bin}_B(\ell)$ and $\mathrm{bin}_B(k-1)$ differ, apply a $\mathrm{CNOT}$ gate to $x_{i,j}$ with $\ket{\text{ancilla}}$ as the control qubit. In this way, the duplicate patterns are rewritten into the basis states of $\mathcal{P}_k^{\mathrm{H}}$. In the above example, since $\mathrm{bin}_B(\ell)=[1,0,0]$ and $\mathrm{bin}_B(k-1)=[0,1,0]$, $\mathrm{CNOT}$ gates are applied to $x_{i,0}$ and $x_{i,1}$, as shown in Figure \ref{fig: basis transition} (b).

\begin{figure*}[h]
    \centering
    \includegraphics[clip,scale=0.55]{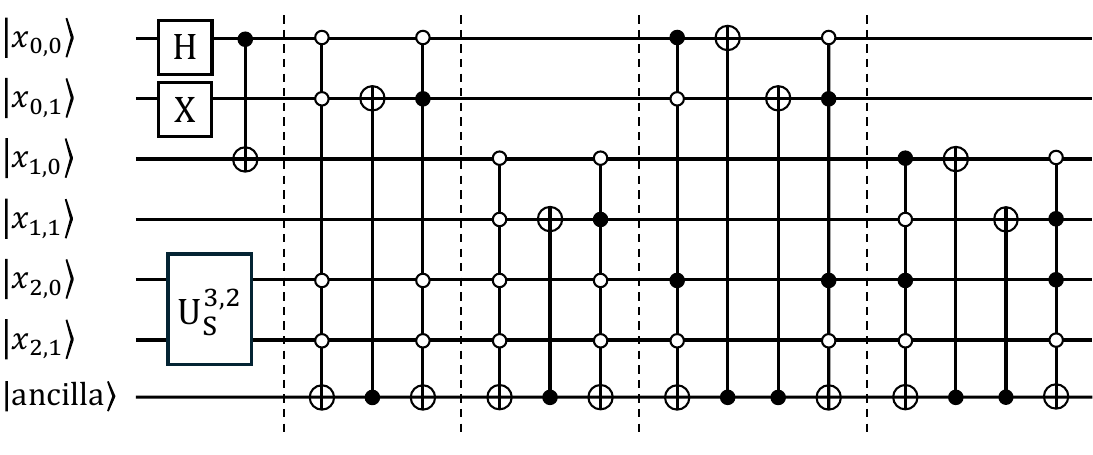}
    \vspace{-4ex}
    \caption{Construction of $\mathrm{U}_\mathrm{P,H}^3$.\label{fig:PPO_HUBO_N=3}}
    \vspace{-2ex}
\end{figure*}

Finally, when duplicate patterns have been rewritten into the basis states of $\mathcal{P}_k^{\mathrm{H}}$, the state of $\ket{\text{ancilla}}$ is reset by applying a multi-controlled $\mathrm{X}$ gate to $\ket{\text{ancilla}}$ with $[x_{i,0},\cdots,x_{i,B-1}]$ and $[x_{k-1,0},\cdots,x_{k-1,B-1}]$ as control qubits. For each $j$ $(0 \leq j \leq B-1)$, if the $j$-th element of $\mathrm{bin}_B(k-1)$ or $\mathrm{bin}_B(\ell)$ is $0$, the corresponding $x_{i,j}$ and $x_{k-1,j}$ are treated as zero-controls. If the element is $1$, they are treated as one-controls. In the example above, the implementation is shown in Figure \ref{fig: basis transition} (c), while the construction of $\mathrm{U}_{\mathrm{P},\mathrm{H}}^{3}$ for the case $N=3$ is illustrated in Figure \ref{fig:PPO_HUBO_N=3}.

\subsubsection{Integrating PPO for HUBO into GAS}

In order to integrate \ac{PPO} for \ac{HUBO} into \ac{GAS}, the uniform superposition state prepared by the state preparation operator $\A_y$ needs to be replaced. Thus, the circuit construction method in Sec.~\ref{subsubsec: initial state preparation of GAS} is revised as follows.


Start by applying \ac{PPO} for \ac{HUBO} $\mathrm{U}_{\mathrm{P},\mathrm{H}}^{N}$ to $n=N\lceil\log_2 N\rceil$ qubits and a single ancilla qubit in the initial state $\ket{0}_{n+1}$ transforms into $\ket{P_{N}^{\mathrm{H}}}\ket{0}$, with the corresponding transition described by
\begin{equation}
\ket{0}_{n+m+1}\xrightarrow{\mathrm{U}_{\mathrm{P},\mathrm{H}}^{N}} \ket{P_{N}^{\mathrm{H}}}\!\!\underbrace{\ket{0}}_{\text{ancilla}}\!\!\ket{0}_m\! =\!\frac{1}{\sqrt{N!}}\!\!\sum_{\x_{\mathrm{H}}\in\mathcal{P}_N^\mathrm{H}}\!\!\ket{\x_\mathrm{H}}_n\ket{0}\!\ket{0}_m.
\end{equation}

Then, as in the original \ac{GAS}, Hadamard gates are applied to $m$ qubits, yielding the transition
\begin{eqnarray}
\frac{1}{\sqrt{N!}}\sum_{\x_{\mathrm{H}}\in\mathcal{P}_N^\mathrm{H}}\ket{\x_\mathrm{H}}_n\ket{0}\ket{0}_m&&\\
&&\hspace{-20ex}\xrightarrow{\H^{\otimes m}} \frac{1}{\sqrt{2^m N!}}\sum_{\x_{\mathrm{H}}\in\mathcal{P}_N^\mathrm{H}}\sum_{z=0}^{2^m-1}\ket{\x_\mathrm{H}}_n\ket{0}\ket{z}_m.\nonumber
\end{eqnarray}

The remaining steps are the same as those presented in \ref{subsec:GAS}. As a result, the transition yielded by the state preparation operator $\A_y$ is modified from \eqref{eq:Ay} to 
\begin{align}
\A_y\ket{0}_{n+m+1}=\frac{1}{\sqrt{N!}} \sum_{\x_\mathrm{H}\in\mathcal{P}_N^\mathrm{H}}\ket{\x_\mathrm{H}}_n\ket{0}\ket{E_\mathrm{H}(\x_\mathrm{H})-y}_m.
\end{align}

\section{Algebraic and Numerical Analysis}
\label{sec:analysis}

In this section, we compare the conventional \ac{QUBO} formulation and the proposed \ac{HUBO} formulation in terms of various metrics, including the number of terms in the objective function, the number of required qubits, the number of required quantum gates, the circuit depth, and the query complexity.

\subsection{The Number of Terms in the Objective Function}

The operator $\U_\mathrm{G}(\theta)$ in \ac{GAS} corresponds to the terms in the objective function, and the number of such terms is one of the key indicators for analyzing the gate count and depth required to construct the state preparation operator $\A_y$. In this subsection, we analyze the number of objective function terms for both the \ac{QUBO} formulation and the \ac{HUBO} formulation.

\subsubsection{QUBO Formulation}

The objective function \eqref{eq:QUBO_obj_permutation} can be expanded as 
\begin{equation}
    \left< \mathrm{F}, \mathrm{X} \mathrm{D} \mathrm{X}^{\mathrm{T}} \right> = \sum_{i,j,k,\ell} f_{i,j} d_{k,\ell} x_{i,k} x_{j,\ell}.
\end{equation}

Under the general definitions of $\mathrm{F}$ and $\mathrm{D}$ with $\mathrm{diag}(\mathbf{F})=0$ and $\mathrm{diag}(\mathbf{D})=0$, we have
\begin{equation}
    \left< \mathrm{F}, \X \D \X^{\mathrm{T}} \right> 
    = \sum_{\substack{i \neq j \\ k \neq \ell}} f_{i,j}\, d_{k,\ell}\, x_{i,k}\, x_{j,\ell}.
\end{equation}

\begin{figure}[h]
\centering
\subfigure[]{
\includegraphics[clip, width=\columnwidth]{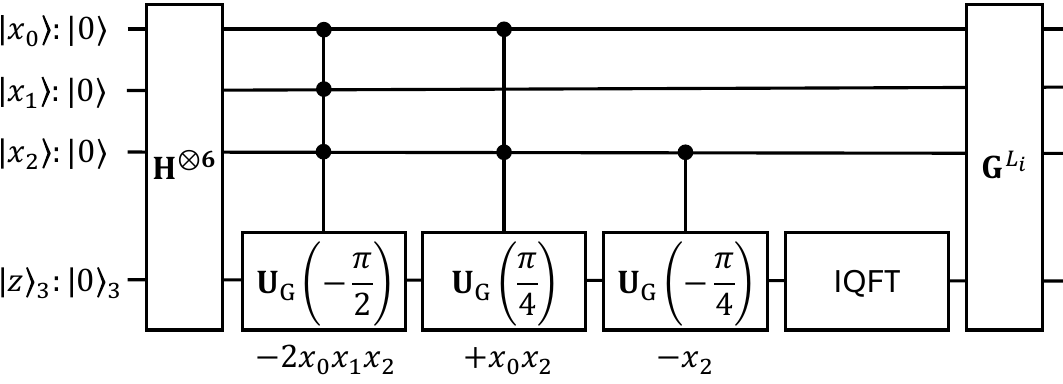}}
\subfigure[]{
\includegraphics[clip, width=\columnwidth]{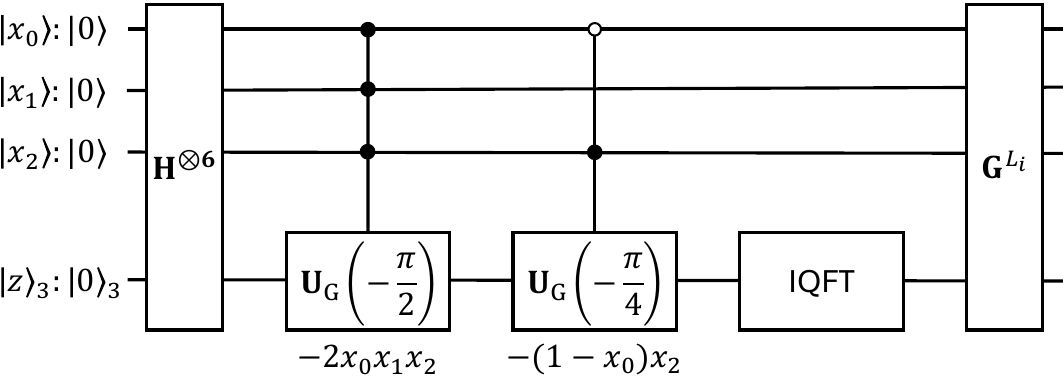}}
\caption{Quantum circuit calculating $E(x)=x_0x_1x_2-(1-x_0)x_2$.}
\label{fig:GAS expansion}
\vspace{-1ex}
\end{figure}

Moreover, since both $\mathrm{F}$ and $\mathrm{D}$ are symmetric matrices, this expression can finally be rewritten as
\begin{equation}
    \left< \mathrm{F}, \mathrm{X} \mathrm{D} \mathrm{X}^{\mathrm{T}} \right> = \sum_{\substack{0\leq i< j\leq N-1\\ k\neq \ell}} 2 f_{i,j} d_{k,\ell} x_{i,k} x_{j,\ell}.
    \label{eq: obj_QUBO_expanded}
\end{equation}

Thus, all terms of the objective function are quadratic, and the total number of terms is $N^2 (N-1)^2 / 2$.

\subsubsection{HUBO Formulation}

The objective function \eqref{eq:HUBO_obj_permutation} can be expanded as 
\begin{equation}
    \left< \mathrm{F}, \mathrm{Y} \mathrm{D} \mathrm{Y}^{\mathrm{T}} \right> = \sum_{\substack{0\leq i< j\leq N-1\\ k\neq \ell}} 2 f_{i,j} d_{k,\ell} y_{i,k} y_{j,\ell}.
    \label{eq: obj_HUBO_expanded}
\end{equation}

Since $y_{i,j}$ is defined by \eqref{eq:delta_ij}, expanding the expression may appear to increase the number of terms in the objective function. However, as shown in \cite{sano2024accelerating}, for an objective function in which only terms of the form $x_{i,j}$ and $(1-x_{i,j})$ appear as in the \ac{HUBO} formulation of this paper the circuit implementation can be carried out without expansion. Specifically, for each $\U_\mathrm{G}(\theta)$ corresponding to a term in the objective function, a one-control is used for $x_{i,j}$, while a zero-control is used for $(1-x_{i,j})$. For example, if the objective function is $E(x)=x_0x_1x_2-(1-x_0)x_2$, then its expansion yields the quantum circuit shown in Figure \ref{fig:GAS expansion} (a). By instead treating the control bit corresponding to $(1-x_0)$ as a zero-control, the implementation becomes Figure \ref{fig:GAS expansion} (b), making it possible to realize the objective function on a quantum circuit without expansion.

Therefore, in the case of the \ac{HUBO} formulation, the number of terms in the objective function can also be regarded as $N^2 (N-1)^2 / 2$, the same as in the \ac{QUBO} formulation.

\subsection{Number of Required Qubits}

The \ac{GAS} algorithm requires $n+m$ qubits for implementation, where $n$ is the number of variables in the objective function and $m$ is the number of bits in the register storing the value of the objective function, satisfying \eqref{eq:constraint_of_m}.

In this section, we analyze the number of qubits required for both the QUBO and HUBO formulations. For simplicity of analysis, we assume that $0 \leq f_{i,j}, d_{k,\ell} \leq 1$.

\subsubsection{QUBO Formulation}

In the \ac{QUBO} formulation, the number of variables is $n = N^2$ for problem size $N$. Next, to determine the number of qubits required to represent the objective function value, we derive a loose upper bound for the objective function. The upper bound of the objective function is given by 
\begin{align}
    \left< \mathrm{F}, \mathrm{X} \mathrm{D} \mathrm{X}^{\mathrm{T}} \right> &= \sum_{i=0}^{N-1}\sum_{j=0}^{N-1} f_{ij}d_{\phi(i)\phi(j)}\nonumber\\
    & \leq \sum_{i,j} f_{i,j}~(\because 0\leq d_{k,\ell}\leq 1)\nonumber\\
    & \leq N^2.
\label{eq:QUBO_upperbound}
\end{align}

\begin{figure}[t]
\centering
\includegraphics[clip, width=\columnwidth]{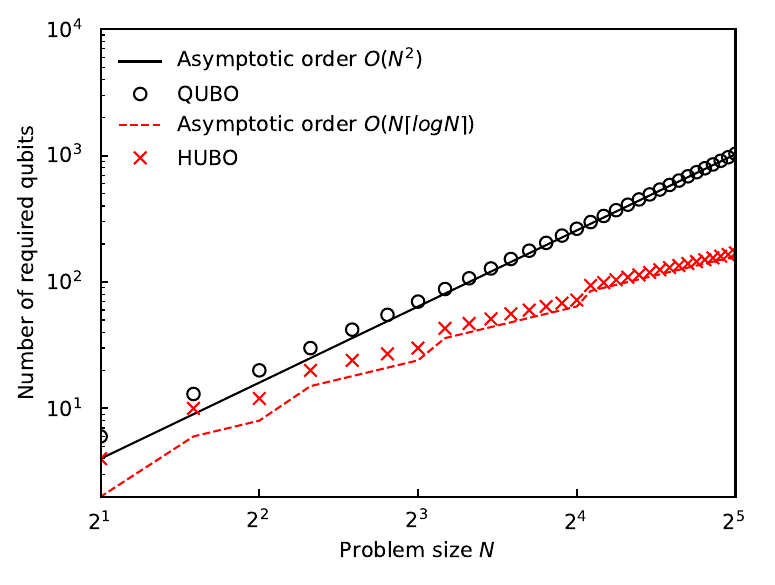}
\caption{Comparison of the number of qubits required for the \ac{QUBO} and \ac{HUBO} formulations.\label{fig:plot_quantumbit}}
\end{figure}

Since the objective value is represented in two’s complement, the required number of qubits is
\begin{equation}
    m = O(\log N).
\end{equation}

Therefore, the total number of qubits needed for the method based on the \ac{QUBO} formulation is
\begin{equation}
    n + m = N^{2} + m = O(N^{2}).
\label{eq:QUBO_qubits}
\end{equation}

\subsubsection{HUBO Formulation}

In the \ac{HUBO} formulation, the number of variables is $n' = N B = N\lceil \log_2 N \rceil$ for problem size $N$, and in the proposed method an additional ancillary qubit is also used. The upper bound of the objective function value can be derived in the same manner as \eqref{eq:QUBO_upperbound},
\begin{equation}
\begin{split}
    \left< \mathrm{F}, \mathrm{Y} \mathrm{D} \mathrm{Y}^{\mathrm{T}} \right> &= \sum_{i=1}^N\sum_{j=1}^N f_{ij}d_{\phi(i)\phi(j)}\leq N^2,
\end{split}
\label{eq:HUBO_upperbound}
\end{equation}
and the number of qubits required to represent the objective function value is
\begin{equation}
    m'=O(\log N).
\end{equation}

Therefore, the total number of qubits required for the method based on the \ac{HUBO} formulation is
\begin{equation}
    n'+m'+1 = N\lceil\log_2 N\rceil + m'+1 = O(N\log N).
\label{eq:HUBO_qubits}
\end{equation}

Figure \ref{fig:plot_quantumbit} compares the number of qubits required for the QUBO and HUBO formulations.
The solid lines indicate the asymptotic order, while the markers represent the actual values calculated from \eqref{eq:QUBO_qubits} and \eqref{eq:HUBO_qubits}. The results show that the HUBO formulation requires fewer qubits, and the advantage becomes more pronounced as the problem size $N$ increases. For small problem sizes, the discrepancies observed between the asymptotic order and the actual values calculated from \eqref{eq:QUBO_qubits} and \eqref{eq:HUBO_qubits} arise because the parameters $m$ and $m'$ cannot be neglected compared with the dominant parameters $n$ and $n'$ in that region.

\subsection{Number of Required Gates}

Following \cite{sano2023qubit,norimoto2023quantum,yukiyoshi2024quantum}, we compare the number of quantum gates required to construct the state preparation operator $\A_y$, which consists of the operator for preparing the initial state, $\U_\mathrm{G}(\theta)$ for computing the objective function value, and the \ac{IQFT} that converts the values computed in the QFT domain into two’s complement representation.

\subsubsection{QUBO Formulation}

First, we analyze the number of quantum gates required to construct the \ac{PPO} for \ac{QUBO} $\U_\mathrm{P,Q}^N$ preparing the initial state. As shown in Figure \ref{fig:QAP_N=2}, $\U_\mathrm{P,Q}^2$ requires one $\X$ gate, one $\H$ gate, and three $\mathrm{CNOT}$ gates. When extending from $\U_\mathrm{P,Q}^{k-1}$ to $\U_\mathrm{P,Q}^{k}$, a W-state preparation operator $\U_\mathrm{W}^k$ and $(k-1)^2$ $\mathrm{CSWAP}$ gates are additionally required. The operator $\U_\mathrm{W}^k$ requires one $\X$ gate, together with $k-1$ $\mathrm{C}R_y$ and $\mathrm{CNOT}$ gates. Taking these into account, the construction of $\U_\mathrm{P,Q}^N$ requires $O(N)$ $\X$ gates, one $\H$ gate, $O(N^2)$ $\mathrm{C}R_y$ gates, $O(N^2)$ $\mathrm{CNOT}$ gates, and $O(N^3)$ $\mathrm{CSWAP}$ gates.

Next, we analyze the number of quantum gates required to construct $\U_\mathrm{G}(\theta)$, which encodes the objective value into the lower $m$ bits register. From \eqref{eq: UGtheta}, each $\U_\mathrm{G}(\theta)$ corresponding to a term of the objective function consists of $m$ phase gates $\R(\theta)$. Moreover, since all terms in the QUBO objective are quadratic and their number is $N^2(N-1)^2/2$, $\U_\mathrm{G}(\theta)$ can be implemented using $O(mN^2)$ $\mathrm{C}^2\R$ gates, where $\mathrm{C}^k\R$ denotes a phase gate with $k$ control bits.

Finally, we analyze the gate count for the \ac{IQFT}. Its implementation requires $m$ Hadamard gates and $m(m-1)/2$ controlled-phase ($\mathrm{C}\R$) gates, where $m$ is the number of bits in the register that stores the \ac{GAS} objective value.

\vspace{-2ex}
\begin{table}[H]
    \caption{Comparison of required quantum gates for $\A_y$ between QUBO and HUBO.}
    \centering
    \begin{tabular}{c|cc}
        \hline
        Quantum gates & QUBO & HUBO\\
        \hline
        $\X$ gates & $O(N)$ & $O(N\log N)$\\
        $\H$ gates & $O(\log N)$ & $O(N)$\\
        $\R_y$ gates & $O(N^2)$ & $O(N)$\\
        $\mathrm{CNOT}$ gates & $O(N^2)$ & $O(N^3\log N)$\\
        $\mathrm{C}\H$ gates & - & $O(N\log N)$\\
        $\mathrm{C}\R_y$ gates & $O(N^2)$ & $O(N\log N)$\\
        $\mathrm{C}\R$ gates & $O(\log^2 N)$ & $O(\log^2 N)$\\
        $\mathrm{C}^2\R$ gates & $O(N^2\log N)$& -\\
        $\mathrm{C}^{2B}\R$ gates & - & $O(N^2\log N)$\\
        $\mathrm{CSWAP}$ gates & $O(N^3)$& -\\
        $\mathrm{C}^{2B}\X$ gates & - & $O(N^3)$\\
        \hline
    \end{tabular}
    \label{tab:required gates}
\end{table}

\subsubsection{HUBO Formulation}

First, we analyze the number of quantum gates required to construct the \ac{PPO} for \ac{HUBO} $\U_\mathrm{P,H}^N$ for preparing the initial state. As shown in Figure \ref{fig:QAP_HUBO_N=2}, $\U_\mathrm{P,Q}^2$ requires one $\X$ gate, one $\H$ gate, and a $\mathrm{CNOT}$ gate. When extending from $\U_\mathrm{P,H}^{k-1}$ to $\U_\mathrm{P,H}^{k}$, the Shukla state preparation operator $\U_\mathrm{S}^{k,B}$ is first applied, followed by the basis-state rewriting depicted in Figure \ref{fig: basis transition} performed $(k-1)^2$ times.

For $\U_\mathrm{S}^{k,B}$, if $k=2^{r}$ $(r\in\mathbb{N})$, it can be implemented with $k$ Hadamard gates. If $k\neq 2^{r}$, write $k=\sum_{i=0}^{p}2^{\ell_i}$ with $0\le \ell_0<\cdots<\ell_p\le B-1$. The required gates are then one $\R_y$ gate, $p$ $\X$ gates, $\ell_0$ $\H$ gates, $(\ell_p-\ell_0)$ controlled-$\H$ gates, and $(p-1)$ controlled-$\R_y$ gates \cite{shukla2024efficient}.

For basis-state rewriting, each qubit string $[x_{i,0},\ldots,x_{i,B-1}]$ is mapped from $\mathrm{bin}_B(\ell)$ to $\mathrm{bin}_B(k)$ $(0\le i,\ell\le k-2)$ using two $\mathrm{C}^{2B}X$ gates and $d_\mathrm{H}(\mathrm{bin}_B(\ell),\mathrm{bin}_B(k))$ CNOTs, where $d_\mathrm{H}(a,b)$ is the Hamming distance for $a=[a_0,\ldots,a_{B-1}]$, $b=[b_0,\ldots,b_{B-1}]$. Thus, $(k-1)^2$ rewrites use $2(k-1)^2$ $\mathrm{C}^{2B}X$ gates and
\begin{equation}
k\sum_{\ell=0}^{k-2}d_\mathrm{H}\big(\mathrm{bin}_B(\ell),\mathrm{bin}_B(k)\big)=O(k^2\log k)
\end{equation}
CNOT gates.

Taking these results into account, the construction of $\U_\mathrm{P,H}^N$ requires $O(N \log N)$ $\X$ gates, $O(N)$ $\H$ gates, $O(N)$ $\R_y$ gates, $O(N^3 \log N)$ $\mathrm{CNOT}$ gates, $O(N \log N)$ controlled-$\H$ gates, $O(N \log N)$ controlled-$\R_y$ gates, and $O(N^3)$ $\mathrm{C}^{2B}\X$ gates.

Next, we analyze the number of quantum gates required to construct $\U_\mathrm{G}(\theta)$. In the \ac{HUBO} formulation, each term in the objective function is of order $2B$, and the total number of terms is $N^2(N-1)^2/2$. Therefore, $\U_\mathrm{G}(\theta)$ can be implemented with $O(m'N^2)$ $\mathrm{C}^{2B}\R$ gates.
For the \ac{IQFT}, as in the \ac{QUBO} case, $m$ Hadamard gates and $m(m-1)/2$ controlled-phase ($\mathrm{C}R$) gates are required.

Table \ref{tab:required gates} presents a comparison of the quantum gate counts required to construct the state preparation operator $\A_y$ for both the \ac{QUBO} and \ac{HUBO} formulations, where we use $m=O(\log N)$ and $m'=O(\log N)$.

\vspace{-1ex}
\subsection{Circuit Depth}

In this subsection, we compare the circuit depth of the state preparation operator $\A_y$ for both the \ac{QUBO} and \ac{HUBO} formulations.

\vspace{-1ex}
\subsubsection{QUBO Formulation}

\begin{figure}[H]
    \centering
    \includegraphics[width=\columnwidth]{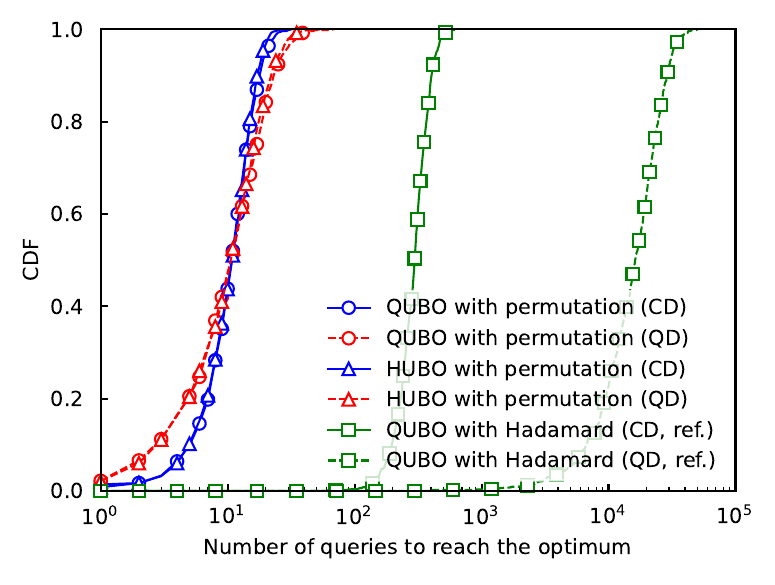}
    \caption{Comparison of query complexity ($N=5$).\label{fig:query}}
\end{figure}

We begin by analyzing the circuit depth of the \ac{PPO} for \ac{QUBO} $\U_\mathrm{P,Q}^N$. For $3 \leq k \leq N$, since $\U_\mathrm{P,Q}^{k-1}$ and $\U_\mathrm{W}^{k}$ are independent, all $\U_\mathrm{W}^{k}$ $(3 \leq k \leq N)$ can be applied in parallel.

Hence, the circuit depth required to generate the W state in each row is $O(\log N)$. Next, consider the depth of $\mathrm{U}_{\mathrm{CSWAP}}^k$. When extending from $\U_\mathrm{P,Q}^{k-1}$ to $\U_\mathrm{P,Q}^{k}$, $\mathrm{U}_{\mathrm{CSWAP}}^k$, consisting of $(k-1)^2$ $\mathrm{CSWAP}$ gates, is required. Since several of these $\mathrm{CSWAP}$ gates can be applied in parallel, the resulting depth is $O(k)$. Since $\mathrm{U}_{\mathrm{CSWAP}}^k$ dominates with depth, the total depth scales as $O(N^2)$. Therefore, the overall depth of $\U_\mathrm{P,Q}^N$ is $O(N^2)$.
Next, the depth of $\U_\mathrm{G}(\theta)$ scales with the number of objective-function terms and, in the worst case, is $m \cdot N^{2}(N-1)^{2}/2 = O(N^{4}\log N)$.
Finally, the depth of the \ac{IQFT} is $m(m+1)/2 = O(\log^{2} N)$.
Combining the above, the depth of $\A_y$ for the QUBO formulation is dominated by $\U_\mathrm{G}(\theta)$ and is $O(N^{4}\log N)$.

\vspace{-1ex}
\subsubsection{HUBO Formulation}

We begin by analyzing the circuit depth of the \ac{PPO} for \ac{HUBO} $\U_\mathrm{P,H}^N$. As in the \ac{QUBO} case, all $\U_\mathrm{S}^{k,B}$ operators that generate Shukla states in each row $(3 \leq k \leq N)$ can be applied simultaneously, resulting in a depth of $O(\log N)$ \cite{shukla2024efficient}. When extending from $\U_\mathrm{P,H}^{k-1}$ to $\U_\mathrm{P,H}^{k}$, it is necessary to apply $(k-1)^2$ groups of gates that rewrite the basis states, as shown in Figure \ref{fig: basis transition}. Since the depth of each gate group is $O(\log k)$, the accumulated depth for $(k-1)^2$ groups becomes $O(k^2 \log k)$. Stacking this over $3 \leq k \leq N$ results in a total depth of $O(N^3 \log N)$.

The depths of $\U_\mathrm{G}(\theta)$ and the \ac{IQFT} are the same as in the \ac{QUBO} case, namely $O(N^4 \log N)$ and $O(\log^2 N)$, respectively.
Therefore, for the \ac{HUBO} formulation as well, the depth of $\A_y$ is dominated by that of $\U_\mathrm{G}(\theta)$, resulting in $O(N^4 \log N)$.

\subsection{Query Complexity}

Finally, we examine the query complexity required when applying the GAS algorithm to solve the \ac{QAP}. As an evaluation metric, we adopt the query complexity in the classical domain (CD) and quantum domain (QD), which is also used in \cite{botsinis2014fixedcomplexity}, defined as the number of measurements of the quantum state and the total number of applications of the Grover operator $\G$, respectively. The analysis considers employing a \ac{PPO} for \ac{QUBO} and \ac{HUBO} formulation, and, for reference, also presents the \ac{QUBO} formulation that prepares the initial state via the Hadamard transform.

Figure \ref{fig:query} shows the cumulative distribution of query complexity for $N=5$. The further to the left the curve lies, the faster the search reaches the optimal solution and the better the convergence performance. For both \ac{QUBO} and \ac{HUBO}, the use of a \ac{PPO} restricts the search space to size $N!$, which is in stark contrast to the Hadamard-based approach whose search space is $2^{N^2}$. The results also indicate that the proposed method achieves performance comparable to that of the conventional approach, and the gap increases as the problem size $N$ grows.

\section{Conclusions}
\label{sec:conc}

We have proposed two novel formulations to solve the \ac{QAP} using the \ac{GAS} algorithm, namely, a \ac{QUBO} formulation and a \ac{HUBO} formulation with \ac{PPO}.
The number of required qubits is $O(N^2)$ for \ac{QUBO} and $O(N\log N)$ for \ac{HUBO}, respectively.
It has been shown that both \ac{QUBO} and \ac{HUBO} significantly improve convergence performance compared with the conventional Hadamard transform. Moreover, although the new PPO for \ac{HUBO} alone results in larger gate counts and depth than the PPO for \ac{QUBO}, from the perspective of the state preparation operator $\A_y$ this overhead is absorbed, yielding comparable depth. As a result, although the depths are comparable, the \ac{HUBO} formulation is superior because it requires fewer qubits.

\footnotesize{
	\bibliographystyle{IEEEtranURLandMonthDiactivated}
	\bibliography{main}
}

\end{document}